\documentclass[aps,preprint,tightenlines,superscriptaddress,showpacs,byrevtex]{revtex4}
\usepackage{graphicx} 
\usepackage{dcolumn}  
\usepackage{amscd}
\usepackage{amsmath}

\graphicspath{{eps/}} 

\begin{document}


\vskip 2cm

\title{ \quad\\[0.5cm]  Measurement of $B \rightarrow D_s^{(*)} K\pi$ branching fractions}

\affiliation{Budker Institute of Nuclear Physics, Novosibirsk}
\affiliation{Chiba University, Chiba}
\affiliation{University of Cincinnati, Cincinnati, Ohio 45221}
\affiliation{T. Ko\'{s}ciuszko Cracow University of Technology, Krakow}
\affiliation{The Graduate University for Advanced Studies, Hayama}
\affiliation{Hanyang University, Seoul}
\affiliation{University of Hawaii, Honolulu, Hawaii 96822}
\affiliation{High Energy Accelerator Research Organization (KEK), Tsukuba}
\affiliation{Hiroshima Institute of Technology, Hiroshima}
\affiliation{Institute of High Energy Physics, Chinese Academy of Sciences, Beijing}
\affiliation{Institute of High Energy Physics, Vienna}
\affiliation{Institute of High Energy Physics, Protvino}
\affiliation{Institute for Theoretical and Experimental Physics, Moscow}
\affiliation{J. Stefan Institute, Ljubljana}
\affiliation{Kanagawa University, Yokohama}
\affiliation{Korea University, Seoul}
\affiliation{Kyungpook National University, Taegu}
\affiliation{\'Ecole Polytechnique F\'ed\'erale de Lausanne (EPFL), Lausanne}
\affiliation{Faculty of Mathematics and Physics, University of Ljubljana, Ljubljana}
\affiliation{University of Maribor, Maribor}
\affiliation{University of Melbourne, School of Physics, Victoria 3010}
\affiliation{Nagoya University, Nagoya}
\affiliation{Nara Women's University, Nara}
\affiliation{National Central University, Chung-li}
\affiliation{National United University, Miao Li}
\affiliation{Department of Physics, National Taiwan University, Taipei}
\affiliation{H. Niewodniczanski Institute of Nuclear Physics, Krakow}
\affiliation{Nippon Dental University, Niigata}
\affiliation{Niigata University, Niigata}
\affiliation{University of Nova Gorica, Nova Gorica}
\affiliation{Osaka City University, Osaka}
\affiliation{Panjab University, Chandigarh}
\affiliation{Saga University, Saga}
\affiliation{University of Science and Technology of China, Hefei}
\affiliation{Seoul National University, Seoul}
\affiliation{Sungkyunkwan University, Suwon}
\affiliation{University of Sydney, Sydney, New South Wales}
\affiliation{Toho University, Funabashi}
\affiliation{Tohoku Gakuin University, Tagajo}
\affiliation{Department of Physics, University of Tokyo, Tokyo}
\affiliation{Tokyo Metropolitan University, Tokyo}
\affiliation{Tokyo University of Agriculture and Technology, Tokyo}
\affiliation{IPNAS, Virginia Polytechnic Institute and State University, Blacksburg, Virginia 24061}
\affiliation{Yonsei University, Seoul}
  \author{J.~Wiechczynski}\affiliation{H. Niewodniczanski Institute of Nuclear Physics, Krakow} 
  \author{T.~Lesiak}\affiliation{H. Niewodniczanski Institute of Nuclear Physics, Krakow}\affiliation{T. Ko\'{s}ciuszko Cracow University of Technology, Krakow} 
  \author{H.~Aihara}\affiliation{Department of Physics, University of Tokyo, Tokyo} 
  \author{K.~Arinstein}\affiliation{Budker Institute of Nuclear Physics, Novosibirsk} 
  \author{V.~Aulchenko}\affiliation{Budker Institute of Nuclear Physics, Novosibirsk} 
  \author{A.~M.~Bakich}\affiliation{University of Sydney, Sydney, New South Wales} 
  \author{V.~Balagura}\affiliation{Institute for Theoretical and Experimental Physics, Moscow} 
  \author{E.~Barberio}\affiliation{University of Melbourne, School of Physics, Victoria 3010} 
  \author{A.~Bay}\affiliation{\'Ecole Polytechnique F\'ed\'erale de Lausanne (EPFL), Lausanne} 
  \author{K.~Belous}\affiliation{Institute of High Energy Physics, Protvino} 
% ...
\collaboration{The Belle Collaboration}

\begin{abstract}
We report a measurement of the exclusive $B^+$ meson decay to the $D_s^{(*)-} K^+\pi^+$
final state using $657 \times 10^{6} B\overline{B}$ pairs
collected at the $\Upsilon(4S)$ resonance with the Belle detector
at the KEKB asymmetric-energy $e^+e^-$ collider.
We use $D_s^{*-}\to D_s^-\gamma$ and the $D_s^-\to \phi\pi^-$, $\overline{K^{*}}(892)^0 K^-$ and
$K^0_S K^-$ decay modes for $D_s^{(*)}$ reconstruction and measure the 
following branching fractions:
${\cal B}(B^+\to D_s^-K^+\pi^+)=
(1.94^{+0.09}_{-0.08} ({\mathrm stat})^{~+0.20}_{~-0.20} {\mathrm (syst)} \pm
0.17 {\cal(B) })\times 10^{-4}$  and
${\cal B}(B^+\to D_s^{*-}K^+\pi^+)=
(1.47^{+0.15}_{-0.14} ({\mathrm stat})^{~+0.19}_{~-0.19} ({\mathrm syst}) \pm
0.13 {\cal(B) })\times 10^{-4}$. The uncertainties are due to  statistics, experimental 
systematic errors and uncertainties of intermediate  branching fractions, respectively. 
\end{abstract}


\maketitle


{\renewcommand{\thefootnote}{\fnsymbol{footnote}}}
\setcounter{footnote}{0}

\begin{figure}[tbh]
\begin{minipage}[b]{.46\linewidth}
\centering
\setlength{\unitlength}{1mm}
\begin{picture}(95,42)
\put(10,40){\bf (a)}
\includegraphics[height=4.5cm,width=8.5cm]{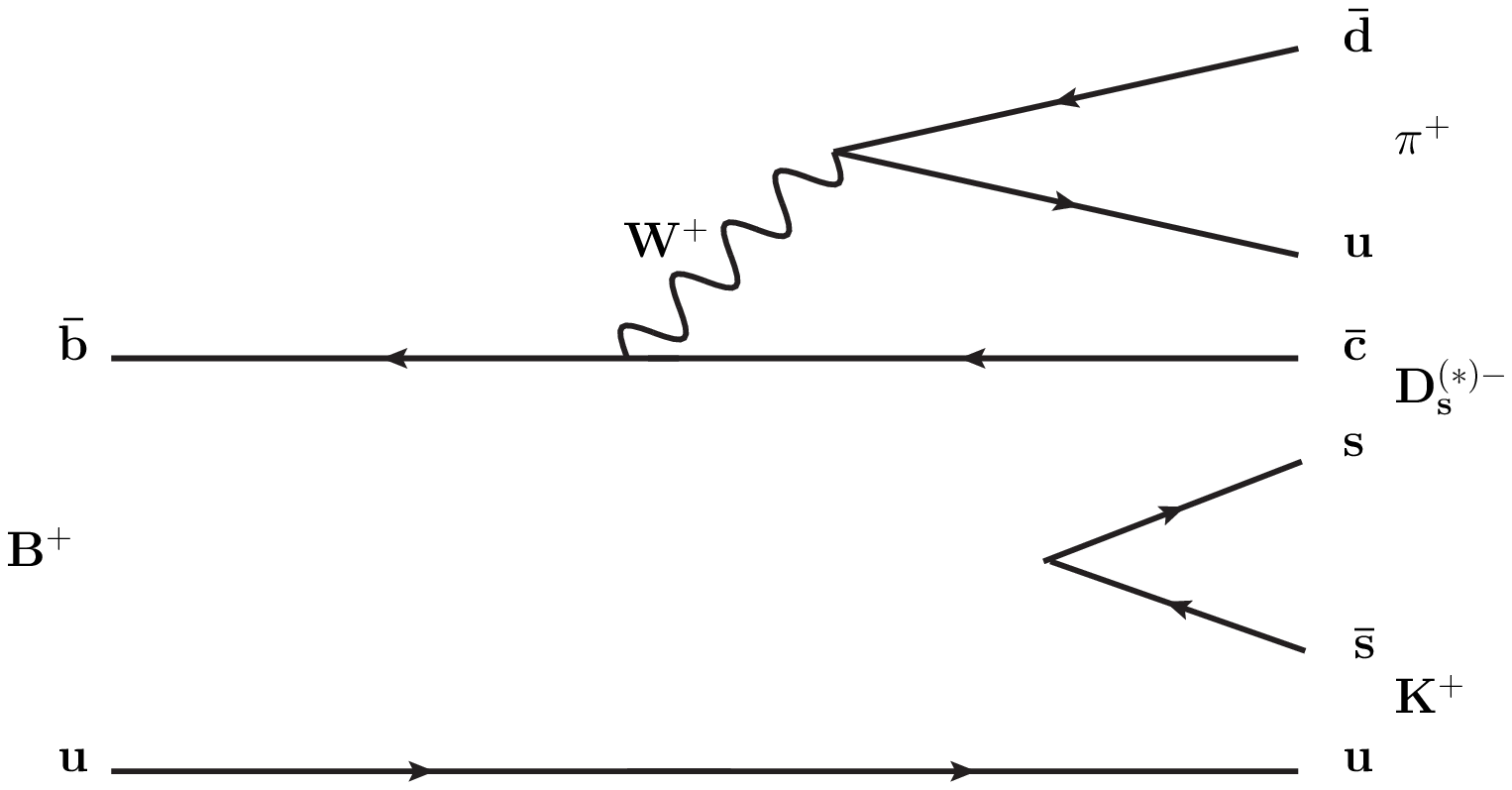}
\end{picture}
\end{minipage}\hfill
\begin{minipage}[b]{.46\linewidth}
\centering
\setlength{\unitlength}{1mm}
\begin{picture}(95,42)
\put(10,40){\bf (b)}
\includegraphics[height=4.5cm,width=8.0cm]{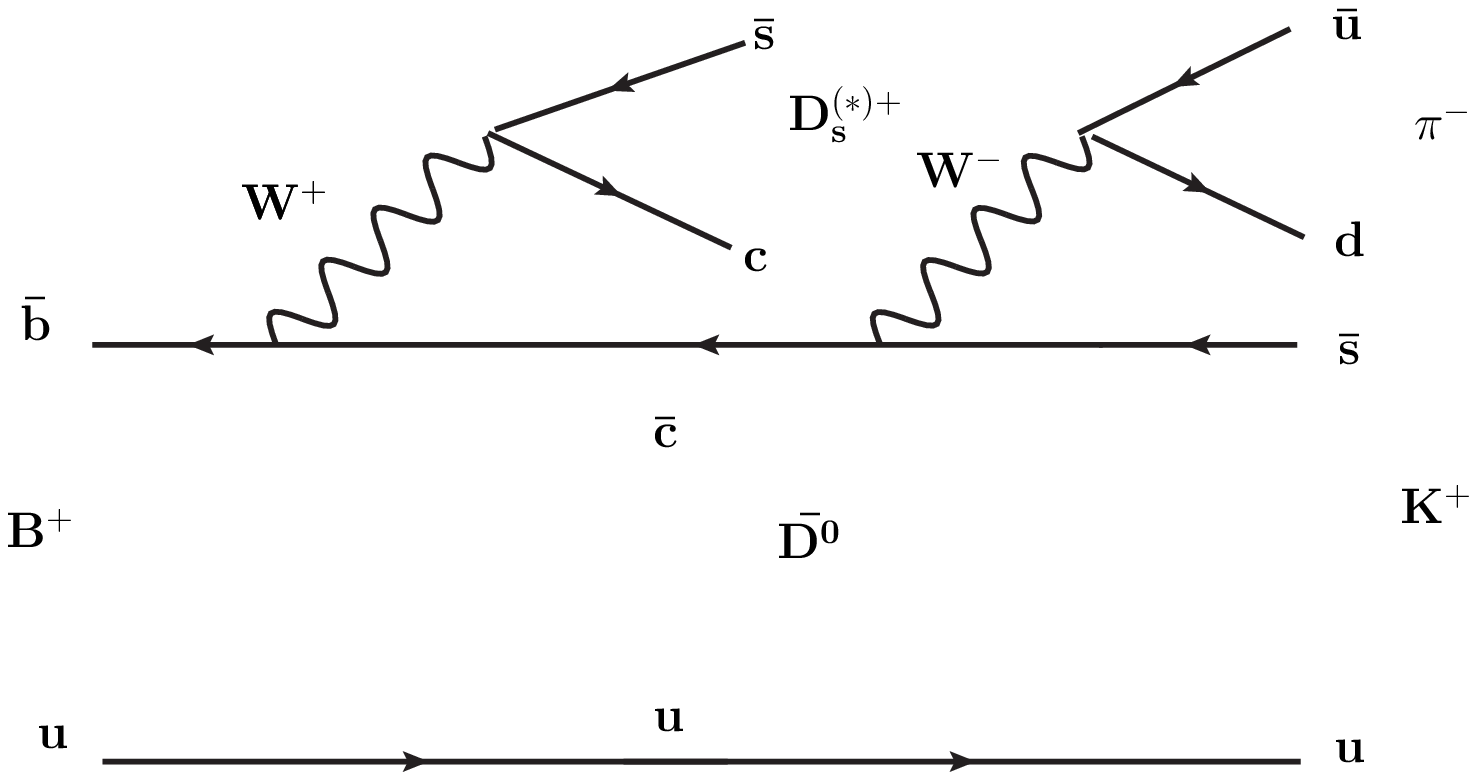}
\end{picture}
\end{minipage}
\caption{ Diagrams for the decays {\bf (a)} $B^+\to D_s^{(*)-} K^+\pi^+$ and {\bf (b)}
$B^+\to D_s^{(*)+} \overline{D^0}, \overline{D^0}\to K^+\pi^-$.
}
\label{FIG1}
\end{figure}

The dominant process in the decays  $B^+\to D_s^{(*)-} K^+\pi^+$~\cite{FOOT} is mediated by the
$b\to c$ quark transition and includes the production of an additional $s\overline{s}$ pair,
 as shown by the Feynman diagram in Fig.~\ref{FIG1}(a).
This process produces at least three final-state particles and can thus be distinguished  from
much more dominant decays, which include direct $D_s$ production from the $W$ boson vertex.
An example of a process that does not involve $s\bar{s}$ quark popping is shown in Fig.~\ref{FIG1}(b); this is the dominant Feynman diagram 
describing two-body $B^+\to D_s^{(*)+}\overline{D}^0$ decays with $\overline{D}^0\to K^+\pi^-$.
Although both $B^+\to D_s^{(*)-} K^+\pi^+$ and $B^+\to D_s^{(*)+}\overline{D}^0 (\overline{D}^0\rightarrow K^+\pi^-)$ decays give a similar  three-body final state, the different decay 
mechanisms lead to opposite charges for the $D_s$ and $\pi$ mesons. In addition, due to the similarities of the final states, the latter decay, $B^+\to D_s^{(*)+}\bar{D}^0$, can be used to check  the experimental procedure
for the exclusive measurements of the former one,  $B^+\to D_s^{(*)-} K^+\pi^+$. These three body decay modes were recently observed by BaBar~\cite{BABAR_DSKAPI} and need further confirmation. 

Studies of $B^+\to D_s^{(*)-} K^+\pi^+$ decays are additionally motivated by interest in the intermediate resonances that can be formed from the three final-state particles. These resonances would be visible in the Dalitz plots
for different two-body subsystems~\cite{THEOR}. 

In this paper we report measurements of the branching fractions 
 for $B^+\to D_s^{(*)-} K^+\pi^+$ decays. We also studied the invariant mass distributions
for the two-body subsystems to search for new resonances. The analysis is performed on a data 
 sample containing $(657\pm 9)\times 10^6$ 
$B\overline{B}$ pairs, collected  
with the  Belle detector at the KEKB asymmetric-energy $e^+e^-$ collider~\cite{KEKB}
that operates at  the $\Upsilon(4S)$ resonance.
The production of $B^+B^-$ and $B^0\overline{B}^0$ pairs is assumed to be equal.


The Belle detector is a large-solid-angle magnetic spectrometer that
consists of a silicon vertex detector (SVD), a 50-layer central drift
chamber (CDC), an array of aerogel threshold Cherenkov counters
(ACC), a barrel-like arrangement of time-of-flight scintillation
counters (TOF), and an electromagnetic calorimeter (ECL) composed of CsI(Tl)
crystals, located inside a superconducting solenoid coil that
provides a 1.5~T magnetic field. An iron flux-return located outside
of the coil is instrumented to detect $K_L^0$ mesons and to identify
muons (KLM). The detector is described in detail
elsewhere~\cite{BELLE}. Two inner detector configurations were
used. A 2.0~cm beam pipe and a 3-layer silicon vertex detector were used
for the first sample of $152 \times 10^6 B\overline{B}$ pairs, while 
a 1.5~cm beam pipe, a 4-layer silicon detector and a small-cell inner drift
chamber were used to record the remaining $505 \times 10^6 B\overline{B}$
pairs \cite{SVD}.


\begin{figure}[htb]


\begin{minipage}[b]{.32\linewidth}
\centering
\setlength{\unitlength}{1mm}
\begin{picture}(60,50)
\put(5,43){\large\bf (a)}
\put(15,-3){\large $\Delta E$[GeV]}
\put(-5,8){\rotatebox{90}{\large\bf $\frac{dN}{d(\Delta E)\;\cdot\;(0.008\;\mathrm{GeV})} $}}
\includegraphics[height=4.5cm,width=5.4cm]{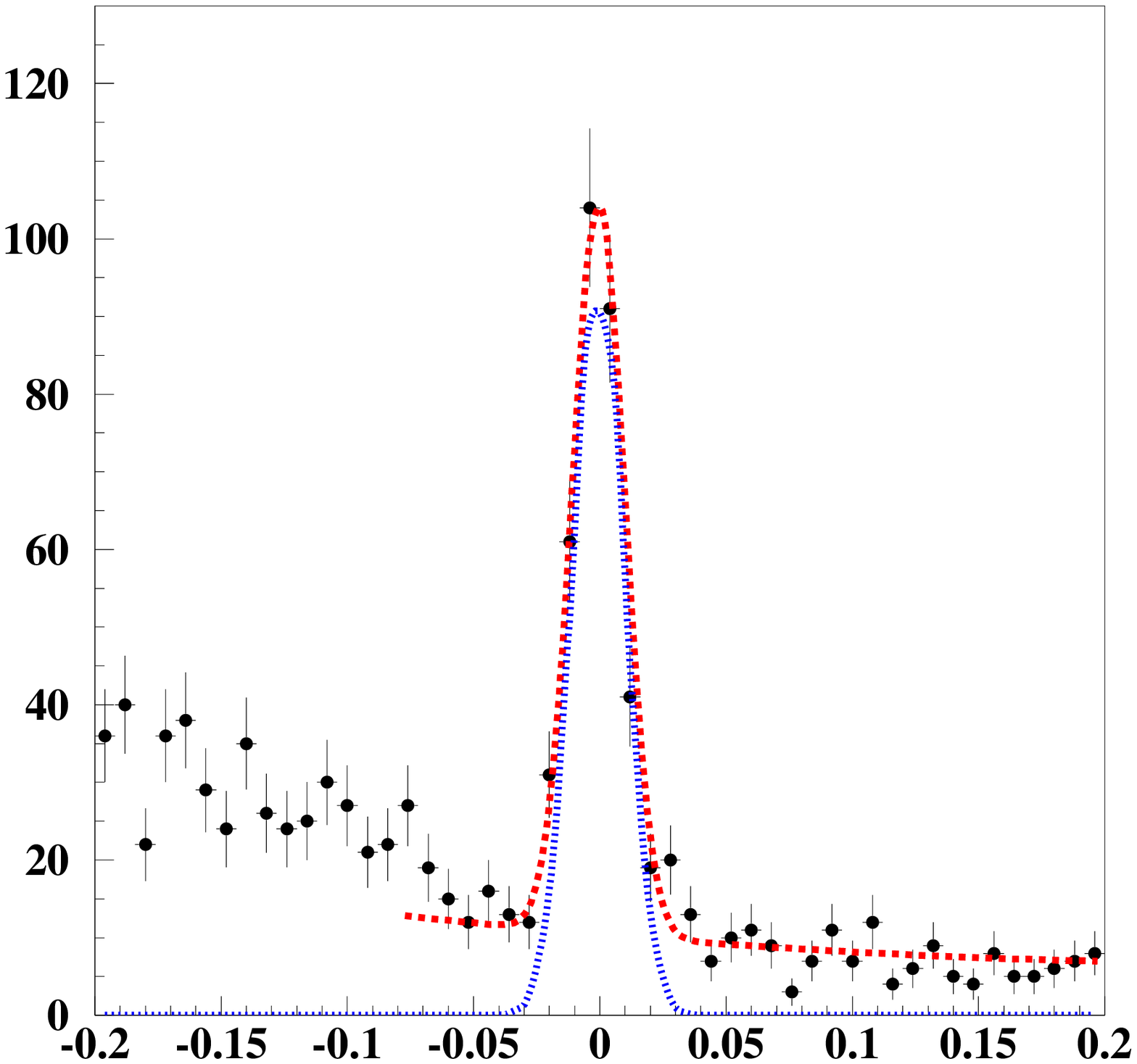}
\end{picture}
\end{minipage}\hfill
\begin{minipage}[b]{.32\linewidth}
\centering
\setlength{\unitlength}{1mm}
\begin{picture}(60,50)
\put(12,-3){\large $M_\mathrm{bc}$[GeV/$c^{2}$]}
\put(-5,4)
{\rotatebox{90}{\large\bf $\frac{dN}{dM_\mathrm{bc}\;\cdot\;(0.002\;\mathrm{GeV}/c^2)}$}}
\includegraphics[height=4.5cm,width=5.4cm]{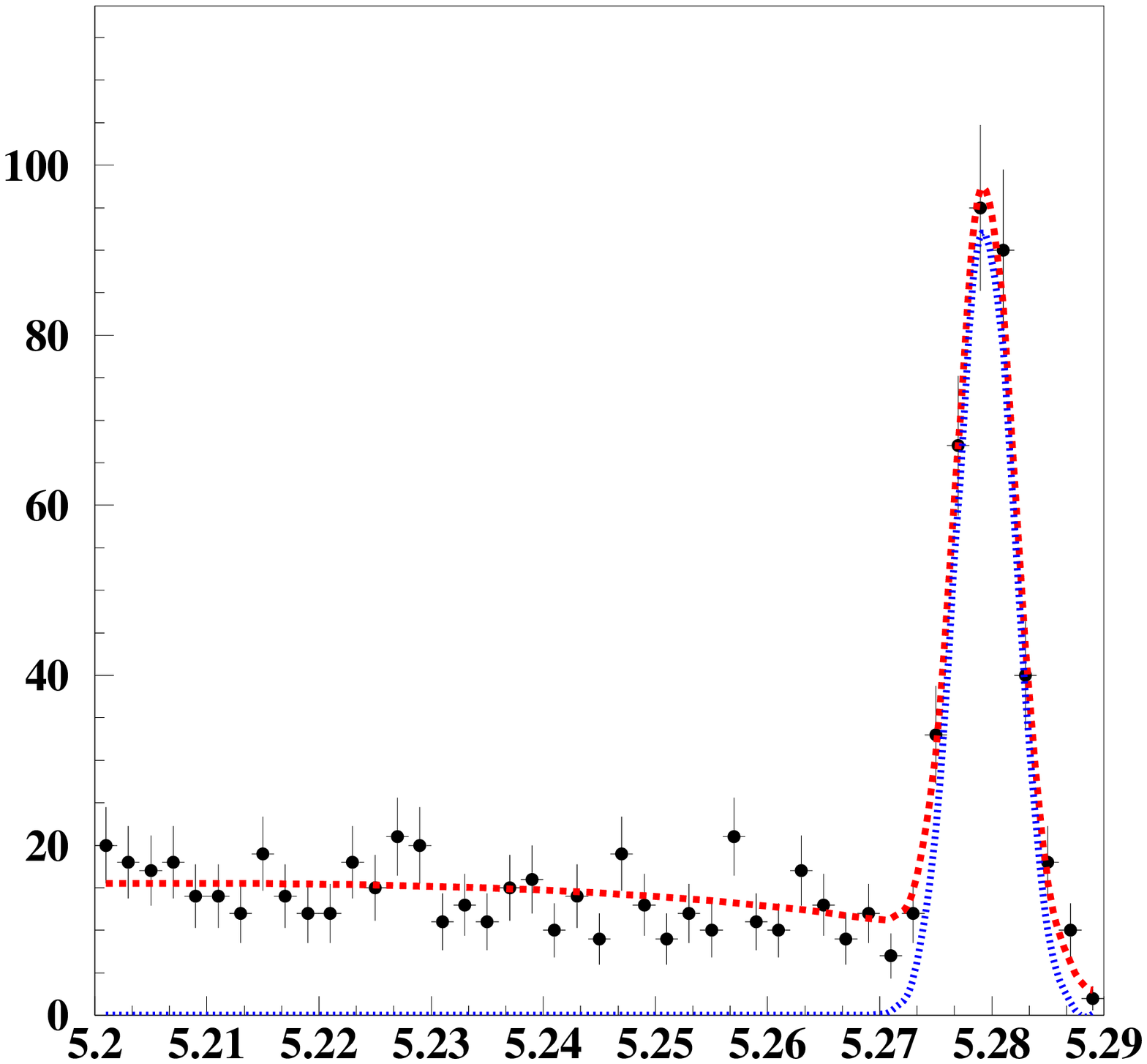}
\end{picture}
\end{minipage}\hfill
\begin{minipage}[b]{.32\linewidth}
\centering
\setlength{\unitlength}{1mm}
\begin{picture}(60,50)
\put(12,-3){\large $M(D_s)$[GeV/$c^{2}$]}
\put(-5,4)
{\rotatebox{90}{\large\bf $\frac{dN}{dM(D_s)\;\cdot\;(0.002\;\mathrm{GeV}/c^2)}$}}
\includegraphics[height=4.5cm,width=5.4cm]{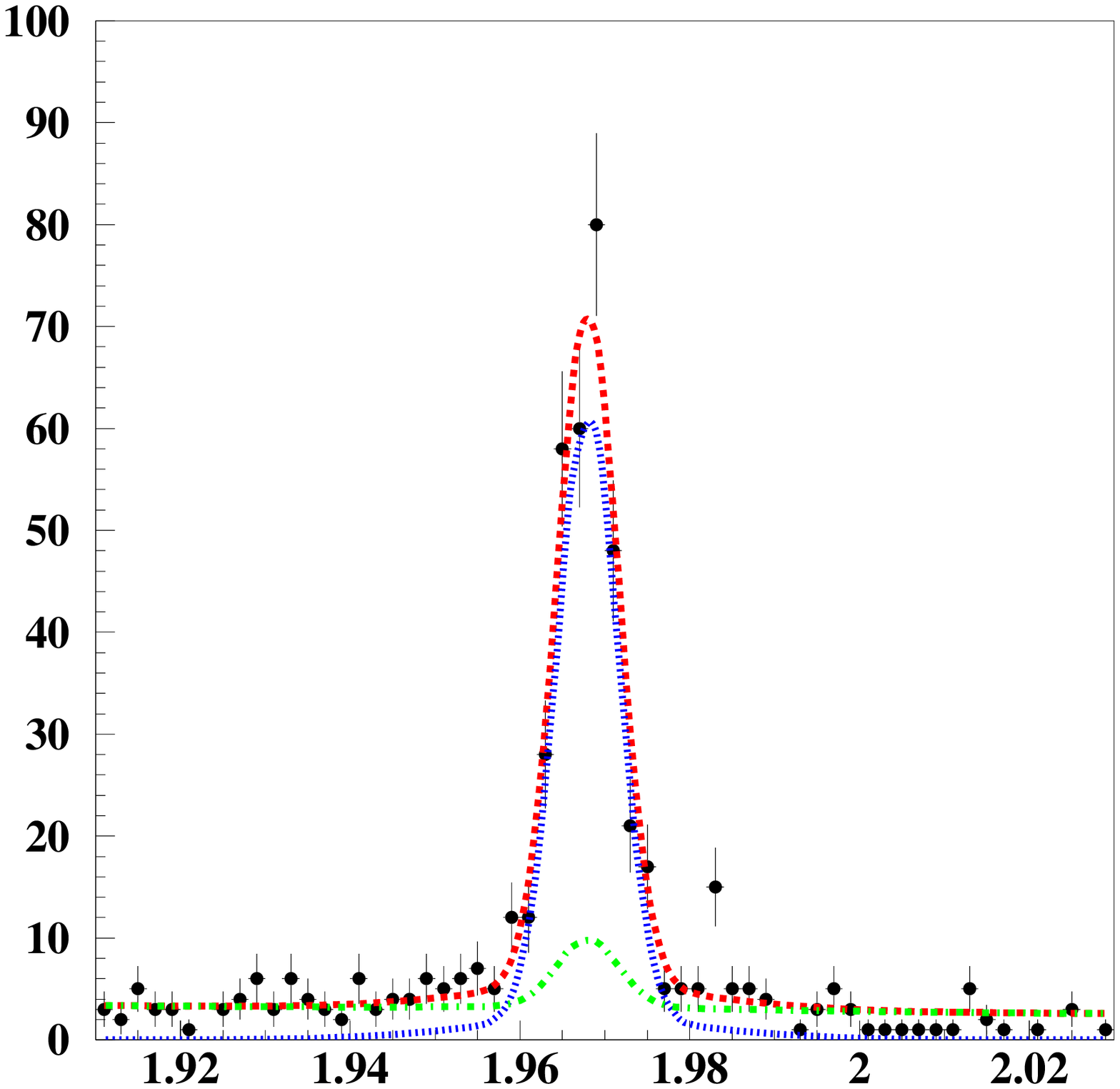}
\end{picture}
\end{minipage}\hfill

\begin{minipage}[b]{.32\linewidth}
\centering
\setlength{\unitlength}{1mm}
\begin{picture}(60,50)
\put(5,43){\large\bf (b)}
\put(15,-3){\large $\Delta E$[GeV]}
\put(-5,8){\rotatebox{90}{\large\bf $\frac{dN}{d(\Delta E)\;\cdot\;(0.008\;\mathrm{GeV})} $}}
\includegraphics[height=4.5cm,width=5.4cm]{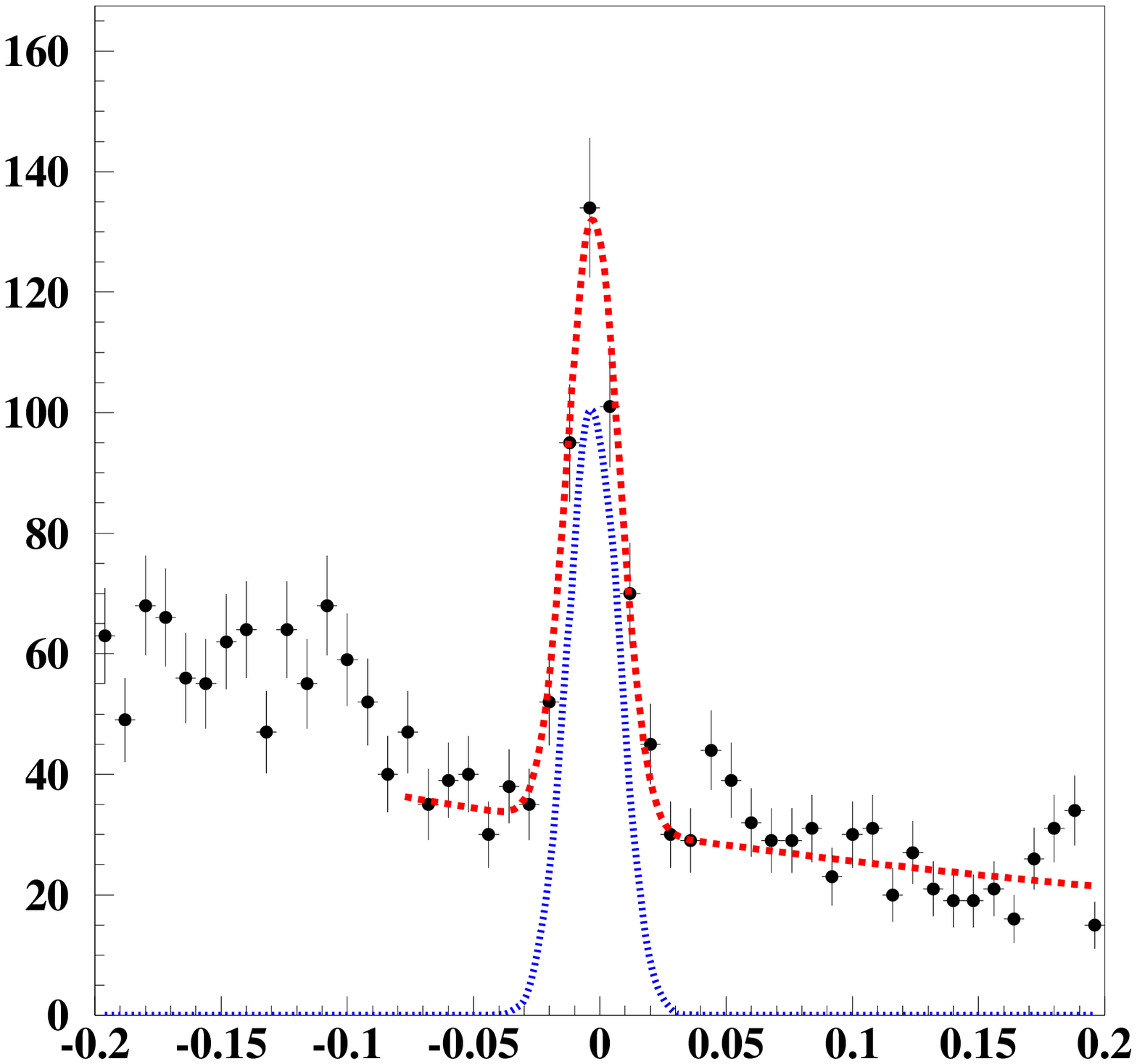}
\end{picture}
\end{minipage}\hfill
\begin{minipage}[b]{.32\linewidth}
\centering
\setlength{\unitlength}{1mm}
\begin{picture}(60,50)
\put(12,-3){\large $M_\mathrm{bc}$[GeV/$c^{2}$]}
\put(-5,4)
{\rotatebox{90}{\large\bf $\frac{dN}{dM_\mathrm{bc}\;\cdot\;(0.002\;\mathrm{GeV}/c^2)}$}}
\includegraphics[height=4.5cm,width=5.4cm]{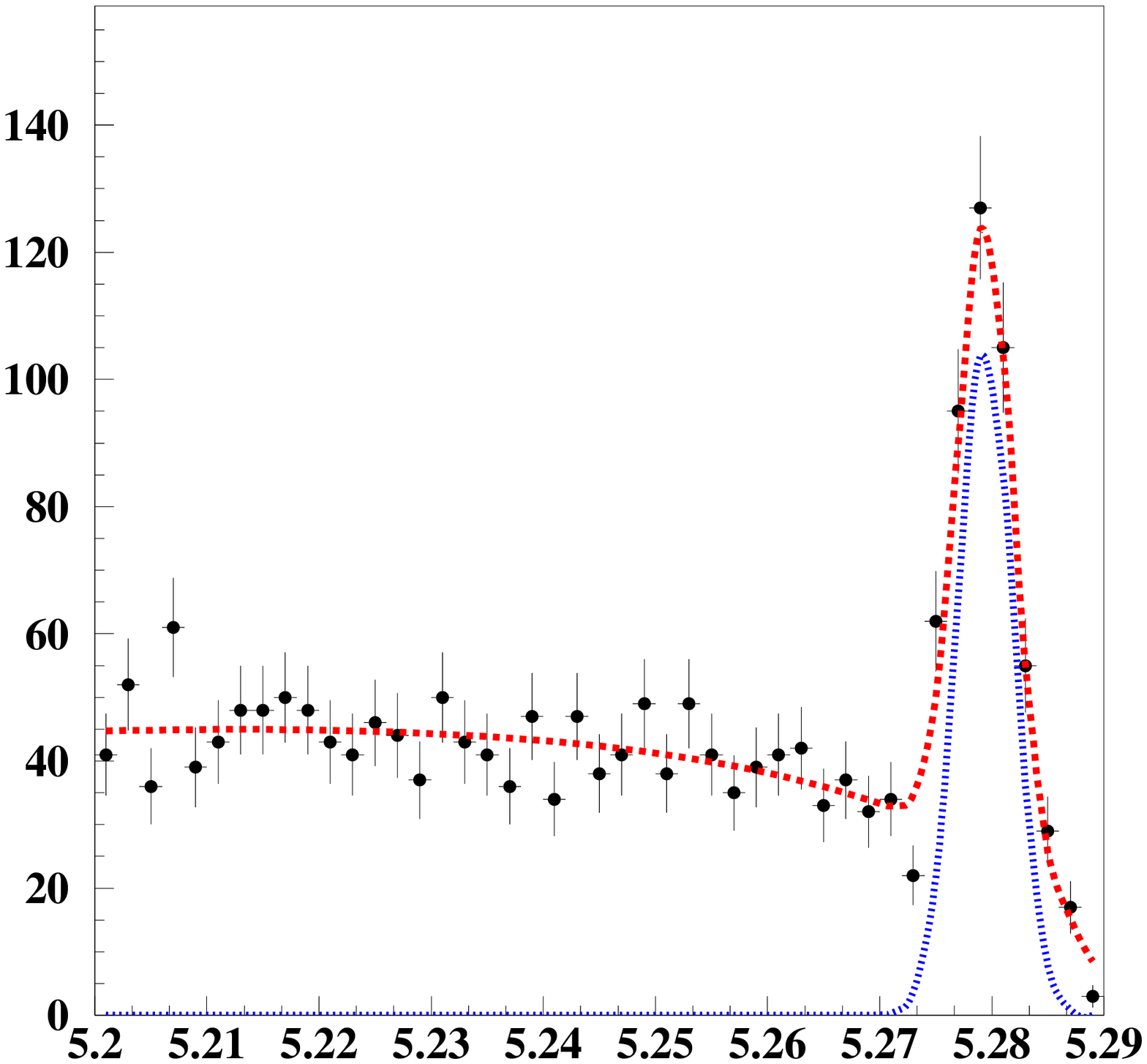}
\end{picture}
\end{minipage}\hfill
\begin{minipage}[b]{.32\linewidth}
\centering
\setlength{\unitlength}{1mm}
\begin{picture}(60,50)
\put(12,-3){\large $M(D_s)$[GeV/$c^{2}$]}
\put(-5,4)
{\rotatebox{90}{\large\bf $\frac{dN}{dM(D_s)\;\cdot\;(0.002\;\mathrm{GeV}/c^2)}$}}
\includegraphics[height=4.5cm,width=5.4cm]{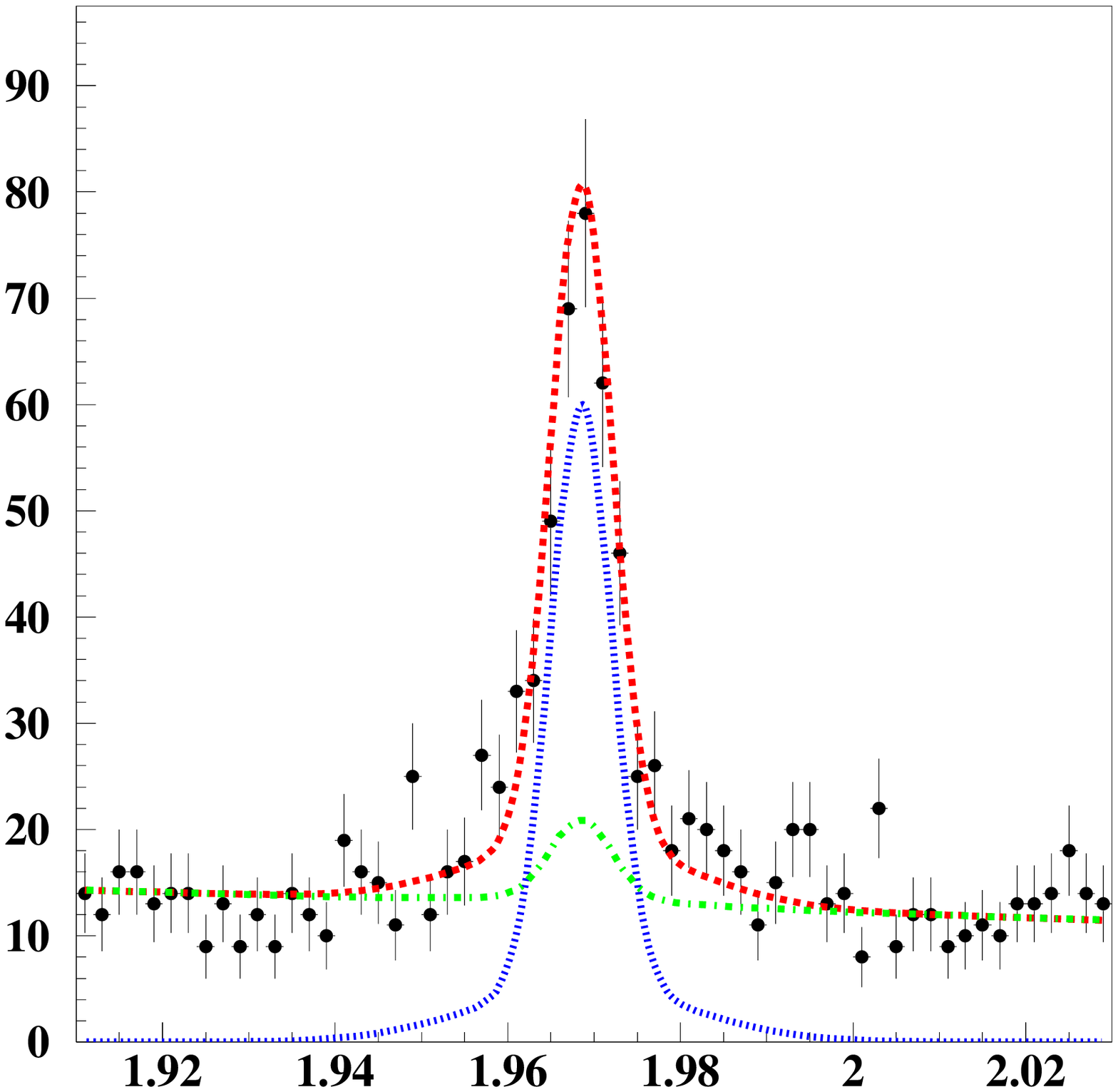}
\end{picture}
\end{minipage}\hfill

\begin{minipage}[b]{.32\linewidth}
\centering
\setlength{\unitlength}{1mm}
\begin{picture}(60,50)
\put(5,43){\large\bf (c)}
\put(15,-3){\large $\Delta E$[GeV]}
\put(-5,8){\rotatebox{90}{\large\bf $\frac{dN}{d(\Delta E)\;\cdot\;(0.008\;\mathrm{GeV})} $}}
\includegraphics[height=4.5cm,width=5.4cm]{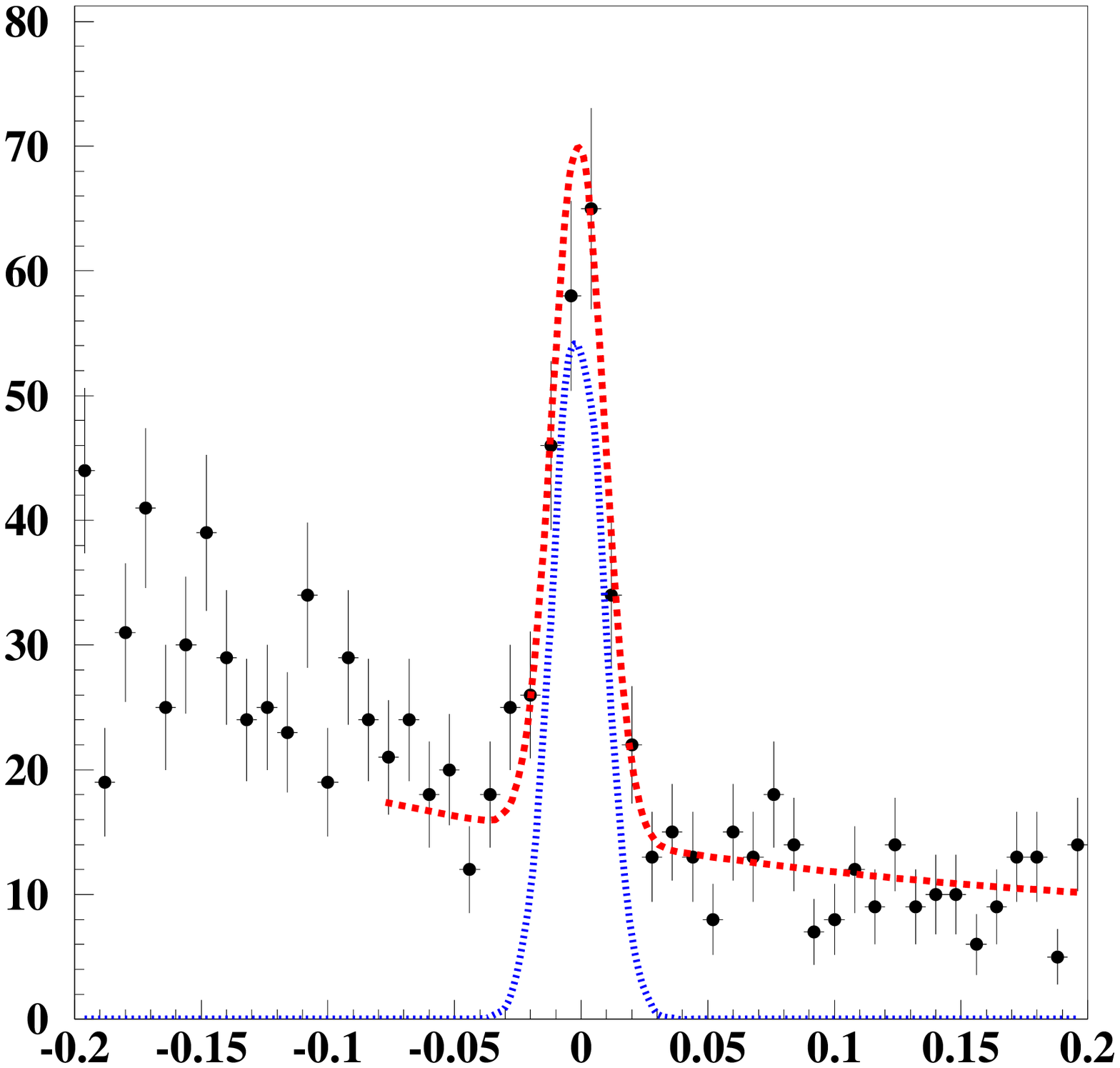}
\end{picture}
\end{minipage}\hfill
\begin{minipage}[b]{.32\linewidth}
\centering
\setlength{\unitlength}{1mm}
\begin{picture}(60,50)
\put(12,-3){\large $M_\mathrm{bc}$[GeV/$c^{2}$]}
\put(-5,4)
{\rotatebox{90}{\large\bf $\frac{dN}{dM_\mathrm{bc}\;\cdot\;(0.002\;\mathrm{GeV}/c^2)}$}}
\includegraphics[height=4.5cm,width=5.4cm]{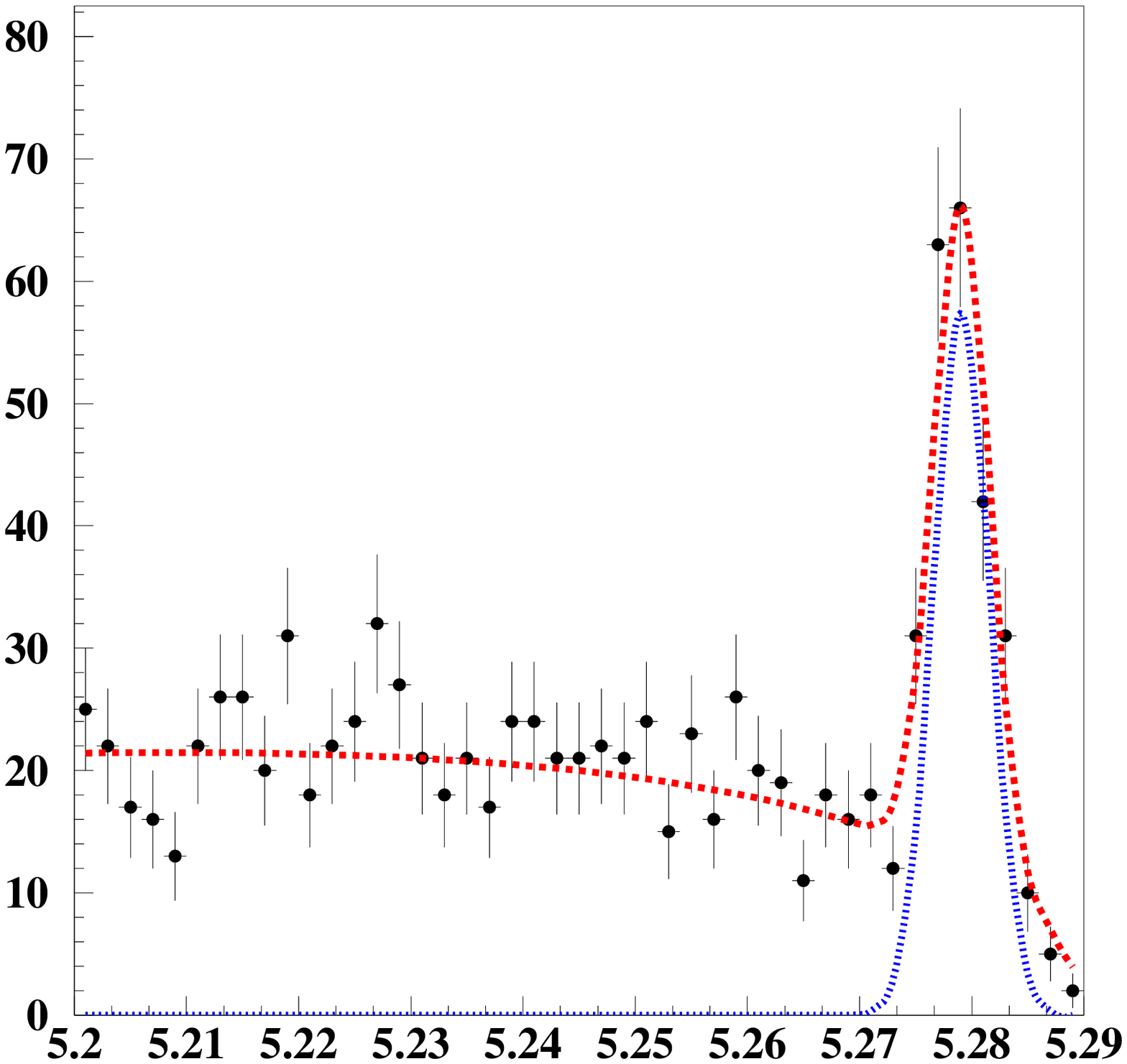}
\end{picture}
\end{minipage}\hfill
\begin{minipage}[b]{.32\linewidth}
\centering
\setlength{\unitlength}{1mm}
\begin{picture}(60,50)
\put(12,-3){\large $M(D_s)$[GeV/$c^{2}$]}
\put(-5,4)
{\rotatebox{90}{\large\bf $\frac{dN}{dM(D_s)\;\cdot\;(0.002\;\mathrm{GeV}/c^2)}$}}
\includegraphics[height=4.5cm,width=5.4cm]{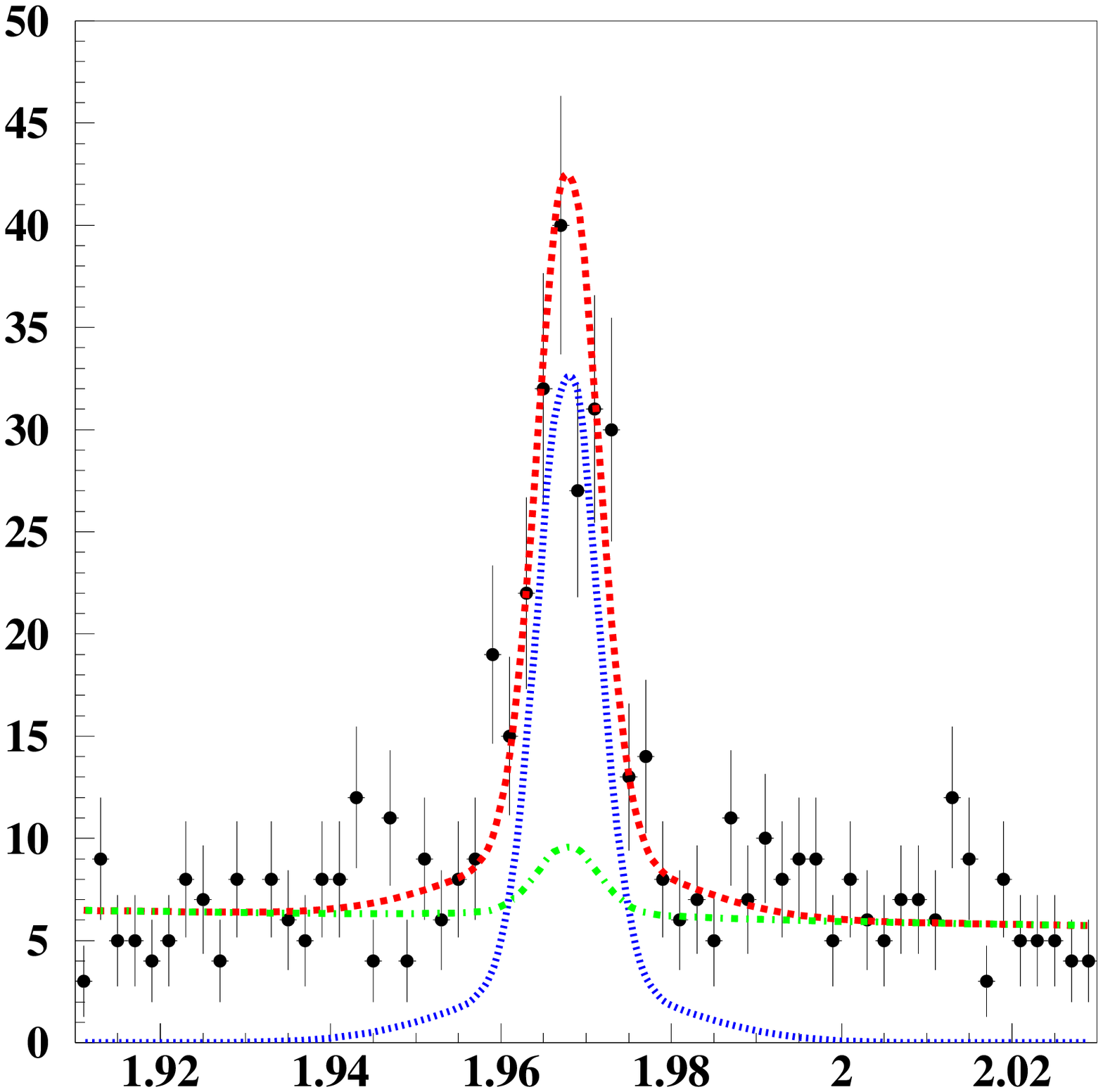}
\end{picture}
\end{minipage}\hfill


\caption{Distributions of $\Delta E$,  $M_\mathrm{bc}$ and $M(D_s) $
for {\bf (a)}  $B^+\to D^{-}_s(\to \phi\pi^-) K^+\pi^+$,
{\bf (b)} $B^+\to D^{-}_s(\to K^{*0} K^-) K^+\pi^+$ and 
{\bf (c)} $B^+\to D^{-}_s(\to K^0_S K^-) K^+\pi^+$ decays.
Each distribution, $\Delta E$,  $M_\mathrm{bc}$ or $M(D_s)$ includes a selection on the signal region of the remaining two.
The red dashed curves show the results of the overall fit described in the text, the blue dashed curves correspond to the signal components and the green dashed curves indicate the fitted background for $M(D_s) $.}

\label{FIG_DSKAPI}
\end{figure}


Charged tracks are required to have a distance of the closest approach 
to the interaction point less than 5~cm in the beam direction (along the $z$-axis)
and less than 5~mm in the transverse ($r-\phi$) plane.
In addition, we only select charged tracks that have transverse momenta larger than 
 $100~{\rm MeV}/c$.

To identify charged hadrons, we combine information from 
 the CDC, ACC and TOF into pion, kaon and proton likelihood variables
${\cal L}_{\pi}$, ${\cal L}_{K}$ and ${\cal L}_{p}$.
 For kaon candidates we then require the likelihood ratio
${\cal L}_{K/\pi} = \frac{{\cal L}_K}{{\cal L}_K + {\cal L}_\pi}$ to be larger than 0.6.
We also apply the proton veto condition: ${\cal L}_{p/K} < 0.95$.
Pions are selected from tracks with low kaon probabilities satisfying a likelihood ratio condition
${\cal L}_{K/\pi} < 0.6$ together with a proton veto
${\cal L}_{p/K} < 0.95$.
In addition, we reject all charged tracks consistent with the electron or muon
hypothesis. The above selection 
results in a typical kaon (pion) identification efficiency
ranging from 92\% to 97\% (94\% to 98\%) for various decay modes,
while 2\% to 15\% of kaon candidates are misidentified pions and
4\% to 8\% of pion candidates are
misidentified kaons.

The $D_s^+$ candidates are reconstructed in three final states:
$\phi(\to K^+K^-)\pi^+ $,  $\overline{K^*}(892)^0(\to K^-\pi^+) K^+ $ and
$K^0_S(\to \pi^+ \pi^-) K^+ $.
 We accept $K^+K^-$  ($K^-\pi^+$) pairs  as $\phi$  ($\overline{K^{*}}(892)^0$) candidates
 if their invariant mass is  within 10 (100) MeV/$c^2$ of the nominal $\phi~( \overline{K^{*}}(892)^0 )$ mass~\cite{PDG}.
This requirement corresponds to $\pm 2.5\sigma$ in all cases. Candidate $K^0_S$ mesons are selected by combining oppositely charged particles with an invariant
mass not differing by more than  6~MeV/$c^2$ from the nominal $K^0_S$ mass. In addition, the vertex of these $\pi^+\pi^-$
pairs  must be displaced from the interaction point by at least 5~mm.
Photons used for $D_s^*\to D_s\gamma$  reconstruction are accepted if their energies exceed 100~MeV
in the laboratory frame. No selection requirements are imposed on the $D_s^{(*)}$ mass at this stage.


A $B$ meson is reconstructed by combining the $D_s$ candidate with
 an identified kaon and pion and by applying
 a loose requirement on the quality ($\chi_B^2$) of the vertex fit to the $K$, $\pi$ and $D_s$ trajectories, where the $D_s$ mass is constrained to the world average
value~\cite{PDG}. The signal $B$ meson decays are  identified 
by three kinematic variables, the $D_s^{(*)}$ invariant mass, the energy difference,
  $\Delta E = E_B - E_\mathrm{beam}$, and the beam-energy-constrained mass,
 $ M_\mathrm{bc} = \sqrt{E_\mathrm{beam}^2 - p_B^2}$.
Here $E_B$ and $p_B$ are the reconstructed energy and momentum of the
$B$ candidate, and $E_\mathrm{beam}$ is the run-dependent beam energy,
all are calculated in the center-of-mass (CM) frame.
For further analysis we retain events in the candidate region defined as:
 $1.91~{\rm GeV/}c^2 < M(D_s) < 2.03$ GeV/$c^2$
 ($2.06~{\rm GeV/}c^2 < M(D_s^*) < 2.16$ GeV/$c^2$),
 $5.2~{\rm GeV/}c^2 < M_\mathrm{bc} < 5.3$ GeV/$c^2$
and $ -0.08~{\rm GeV} < \Delta E < 0.2$ GeV. The lower bound in $\Delta E$ 
for candidate events is chosen to exclude a possible background from
$B\to D_s X$ decays with higher multiplicities.
From GEANT~\cite{GEANT} based Monte Carlo (MC) simulation, we deduce that 
the signal peaks in a signal box are defined by the requirements:
 $1.9532~{\rm GeV/}c^2 < M(D_s) < 1.9832$ GeV/$ c^2$
 ($2.092~{\rm GeV/}c^2 < M(D_s^*) < 2.132$ GeV/$c^2$),
 $5.27~{\rm GeV/}c^2 < M_\mathrm{bc} < 5.29$ GeV/$c^2$ and
$|\Delta E| < 0.03$ GeV.
Based on MC simulation, the region  2.88 GeV/$c^2 <
M(K^+K^-\pi^+\pi^-) <$ 3.18 GeV/$c^2$ is excluded
to remove background from 
$B^+\to ((c\overline{c})\to K^+K^-\pi^+\pi^-) K^+$ decays, where
$(c\overline{c})$ are charmonium states such as the $J/\psi$ or $\eta_c$.
For $B^+ \rightarrow D_s^+ \overline{D^0}(\to K^+\pi^-)$ decays the events in the candidate region
are required to have the $K^+\pi^-$ invariant mass within a 15~MeV/$c^2$ ($3\sigma$) interval of the nominal
 $D^0$ mass.

We find that for  $B^+\to D_s^- K^+\pi^+$ ($B^+\to D_s^{*-} K^+\pi^+$)  decays 
 at most   11\% (29\%) of events  have more
than one $B$ candidate. In such cases we select the $B$ candidate
with the smallest value of $\chi_B^2$. Moreover, when there are at least two
combinations with the same  $\chi_B^2$ value, the one containing a
kaon -- originating directly from the $B$ decay -- with the highest
likelihood ratio ${\cal L}_{K/\pi}$ is selected. For $B\to D_s^* K \pi$ decays,
 we further choose the combination
that minimizes the quantity $| M(D_s^*) - M(D_s) - 143.8~{\rm MeV/}c^2|$.

We exploit the event topology to discriminate between spherical
$B\overline{B}$ events and the dominant background from jet-like continuum
events, $e^{+}e^{-} \rightarrow q\overline{q}$ ($q$ = $u$, $d$, $s$,
$c$). We use the event shape variable $R_2$ defined as the ratio
of the second and zeroth  Fox-Wolfram moments~\cite{FOX} 
and require that $R_2$ be less than 0.4.


\begin{figure}[htb]


\begin{minipage}[b]{.32\linewidth}
\centering
\setlength{\unitlength}{1mm}
\begin{picture}(60,50)
\put(5,43){\large\bf (a)}
\put(15,-3){\large $\Delta E$[GeV]}
\put(-5,8){\rotatebox{90}{\large\bf $\frac{dN}{d(\Delta E)\;\cdot\;(0.008\;\mathrm{GeV})} $}}
\includegraphics[height=4.5cm,width=5.4cm]{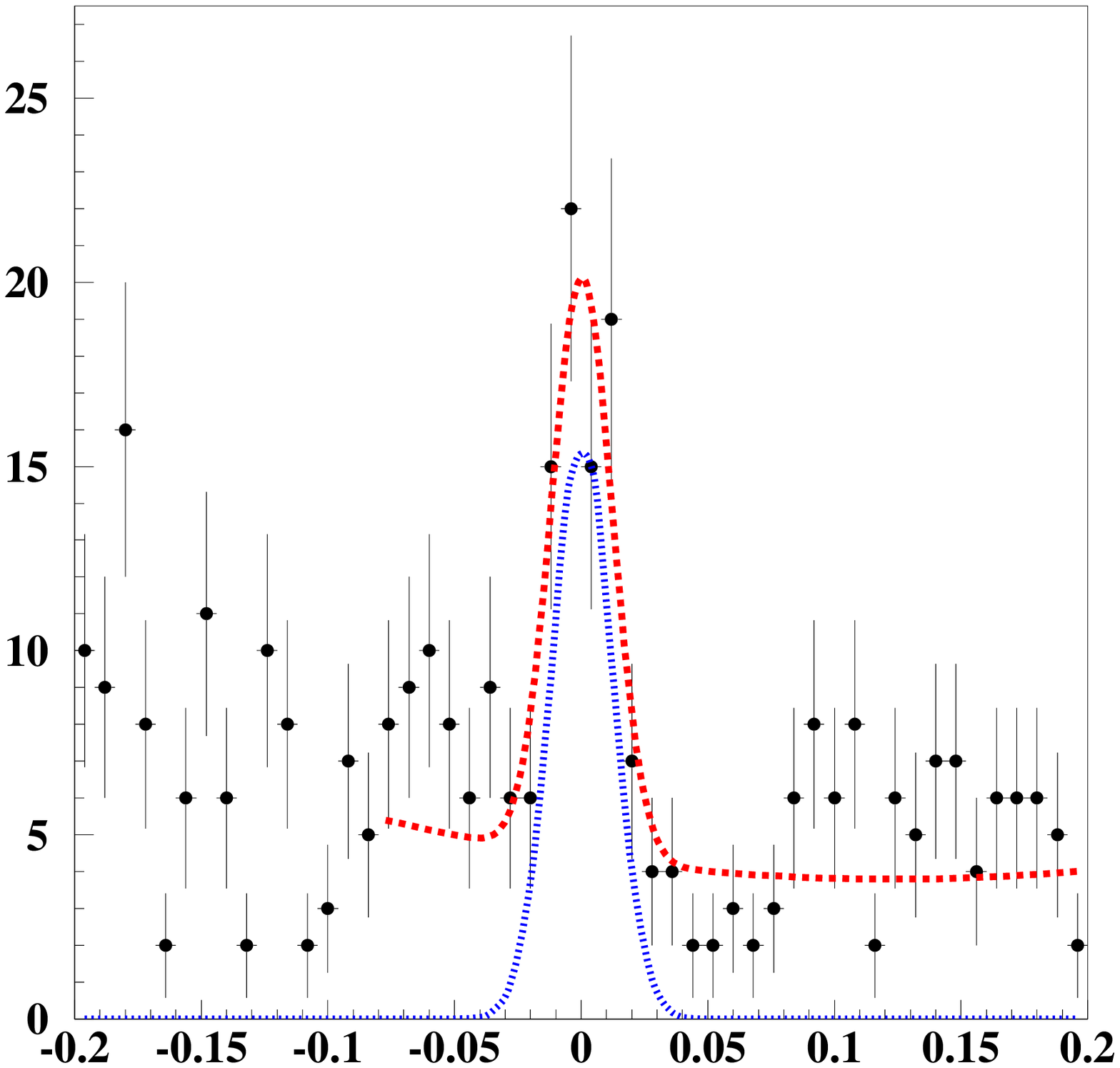}
\end{picture}
\end{minipage}\hfill
\begin{minipage}[b]{.32\linewidth}
\centering
\setlength{\unitlength}{1mm}
\begin{picture}(60,50)
\put(12,-3){\large $M_\mathrm{bc}$[GeV/$c^{2}$]}
\put(-5,4)
{\rotatebox{90}{\large\bf $\frac{dN}{dM_\mathrm{bc}\;\cdot\;(0.002\;\mathrm{GeV}/c^2)}$}}
\includegraphics[height=4.5cm,width=5.4cm]{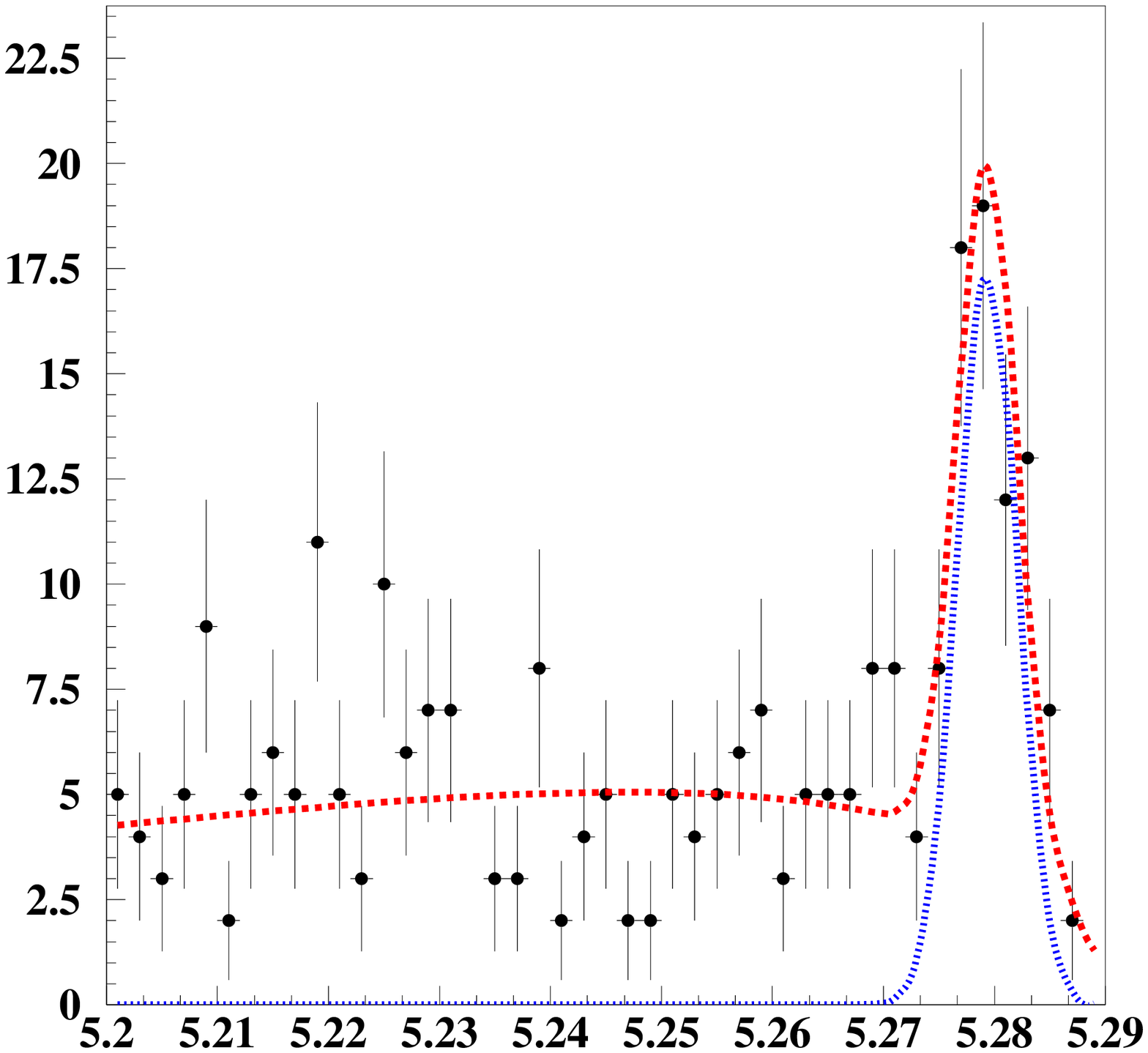}
\end{picture}
\end{minipage}\hfill
\begin{minipage}[b]{.32\linewidth}
\centering
\setlength{\unitlength}{1mm}
\begin{picture}(60,50)
\put(12,-3){\large $M(D_s^*)$[GeV/$c^{2}$]}
\put(-5,4)
{\rotatebox{90}{\large\bf $\frac{dN}{dM(D_s^*)\;\cdot\;(0.002\;\mathrm{GeV}/c^2)}$}}
\includegraphics[height=4.5cm,width=5.4cm]{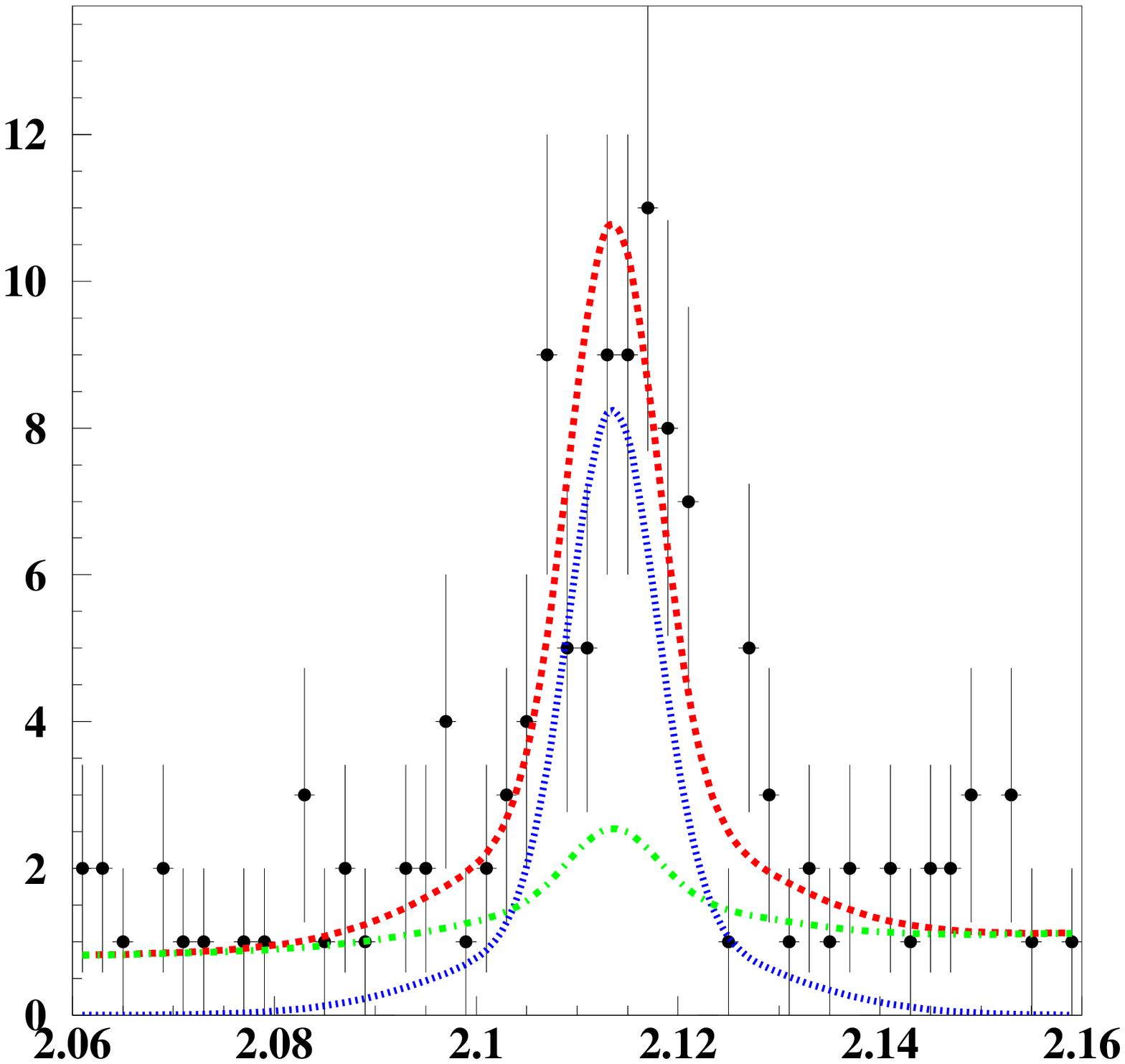}
\end{picture}
\end{minipage}\hfill

\begin{minipage}[b]{.32\linewidth}
\centering
\setlength{\unitlength}{1mm}
\begin{picture}(60,50)
\put(5,43){\large\bf (b)}
\put(15,-3){\large $\Delta E$[GeV]}
\put(-5,8){\rotatebox{90}{\large\bf $\frac{dN}{d(\Delta E)\;\cdot\;(0.008\;\mathrm{GeV})} $}}
\includegraphics[height=4.5cm,width=5.4cm]{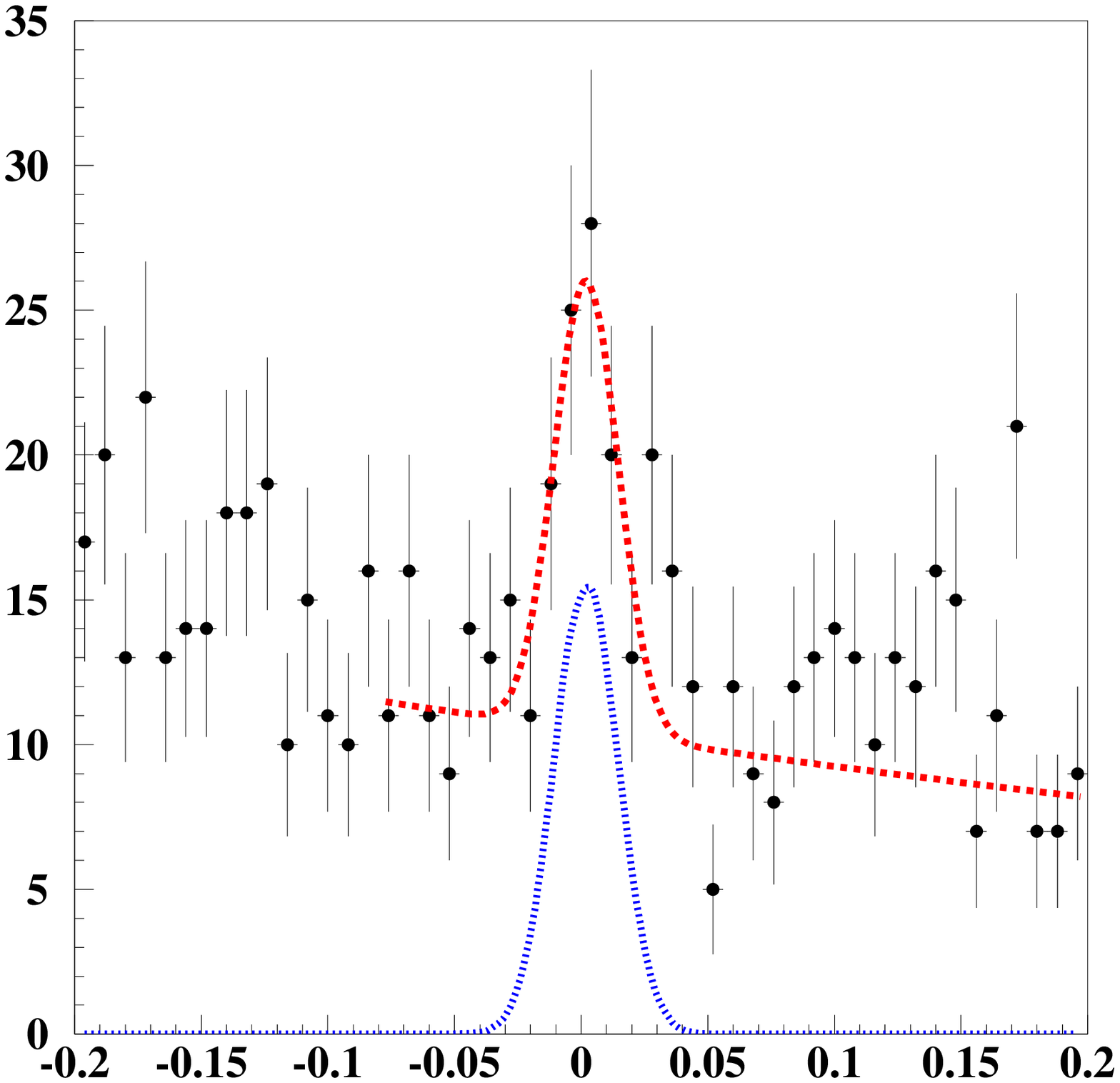}
\end{picture}
\end{minipage}\hfill
\begin{minipage}[b]{.32\linewidth}
\centering
\setlength{\unitlength}{1mm}
\begin{picture}(60,50)
\put(12,-3){\large $M_\mathrm{bc}$[GeV/$c^{2}$]}
\put(-5,4)
{\rotatebox{90}{\large\bf $\frac{dN}{dM_\mathrm{bc}\;\cdot\;(0.002\;\mathrm{GeV}/c^2)}$}}
\includegraphics[height=4.5cm,width=5.4cm]{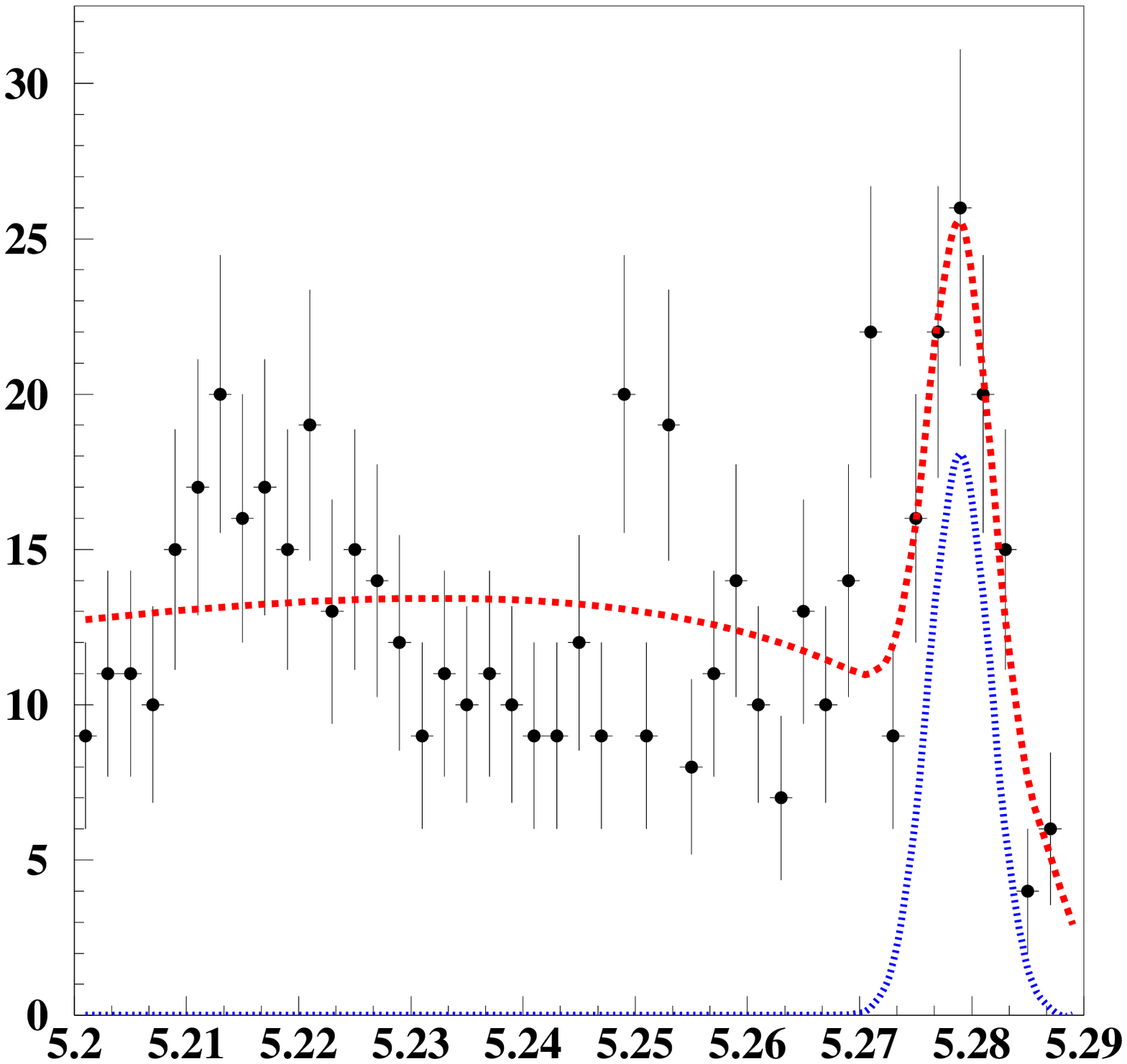}
\end{picture}
\end{minipage}\hfill
\begin{minipage}[b]{.32\linewidth}
\centering
\setlength{\unitlength}{1mm}
\begin{picture}(60,50)
\put(12,-3){\large $M(D_s^*)$[GeV/$c^{2}$]}
\put(-5,4)
{\rotatebox{90}{\large\bf $\frac{dN}{dM(D_s^*)\;\cdot\;(0.002\;\mathrm{GeV}/c^2)}$}}
\includegraphics[height=4.5cm,width=5.4cm]{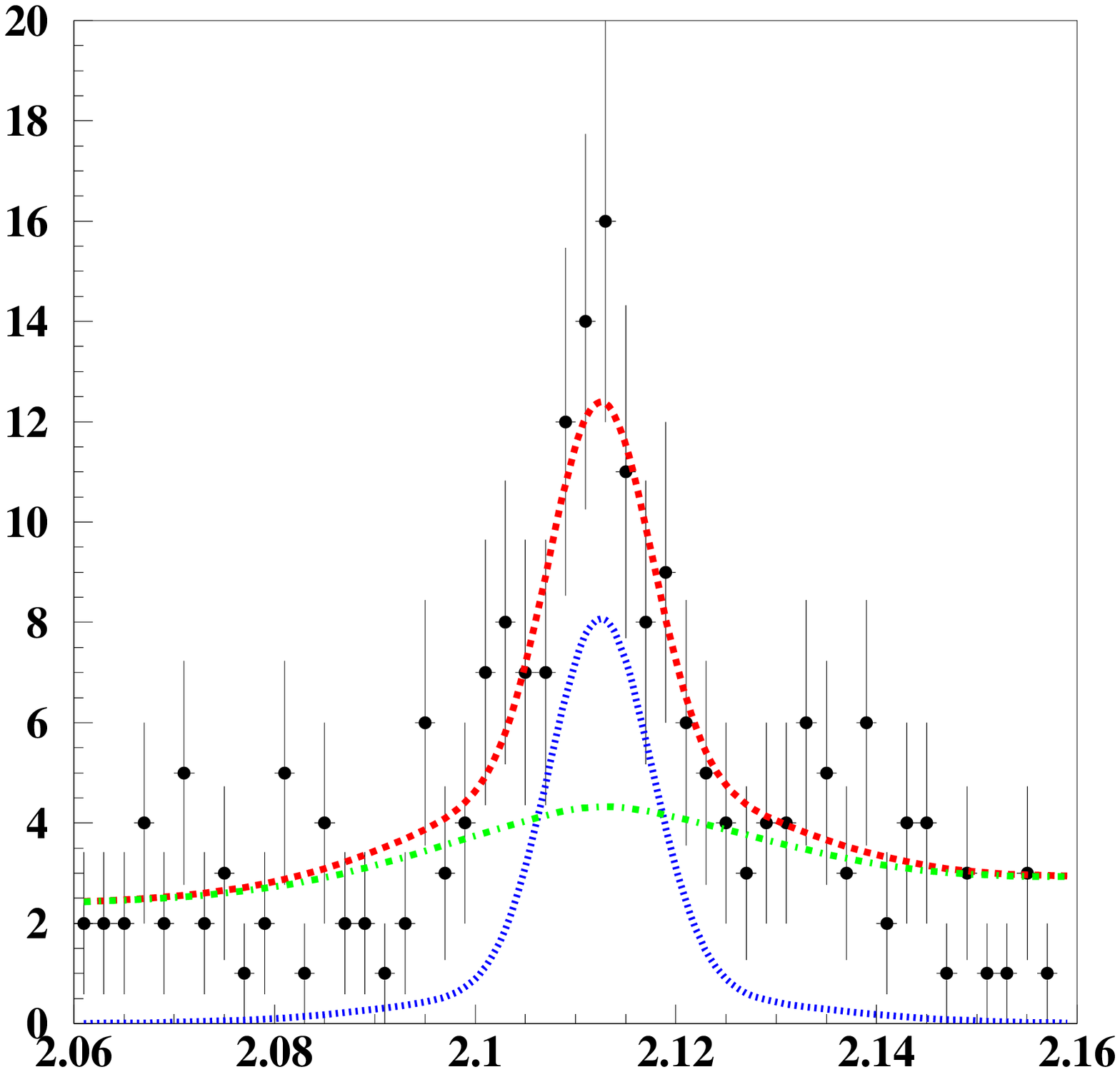}
\end{picture}
\end{minipage}\hfill

\begin{minipage}[b]{.32\linewidth}
\centering
\setlength{\unitlength}{1mm}
\begin{picture}(60,50)
\put(5,43){\large\bf (c)}
\put(15,-3){\large $\Delta E$[GeV]}
\put(-5,8){\rotatebox{90}{\large\bf $\frac{dN}{d(\Delta E)\;\cdot\;(0.008\;\mathrm{GeV})} $}}
\includegraphics[height=4.5cm,width=5.4cm]{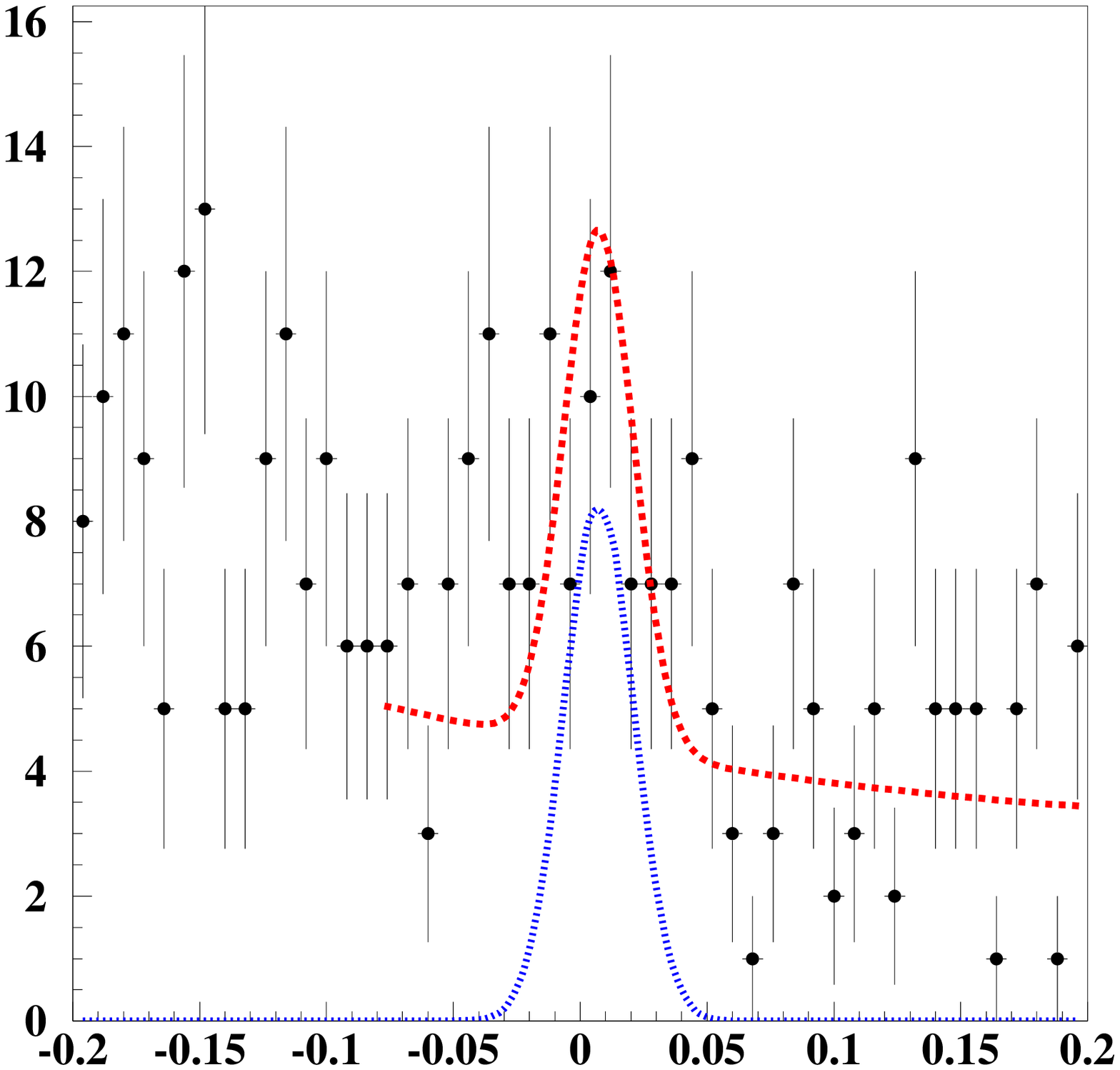}
\end{picture}
\end{minipage}\hfill
\begin{minipage}[b]{.32\linewidth}
\centering
\setlength{\unitlength}{1mm}
\begin{picture}(60,50)
\put(12,-3){\large $M_\mathrm{bc}$[GeV/$c^{2}$]}
\put(-5,4)
{\rotatebox{90}{\large\bf $\frac{dN}{dM_\mathrm{bc}\;\cdot\;(0.002\;\mathrm{GeV}/c^2)}$}}
\includegraphics[height=4.5cm,width=5.4cm]{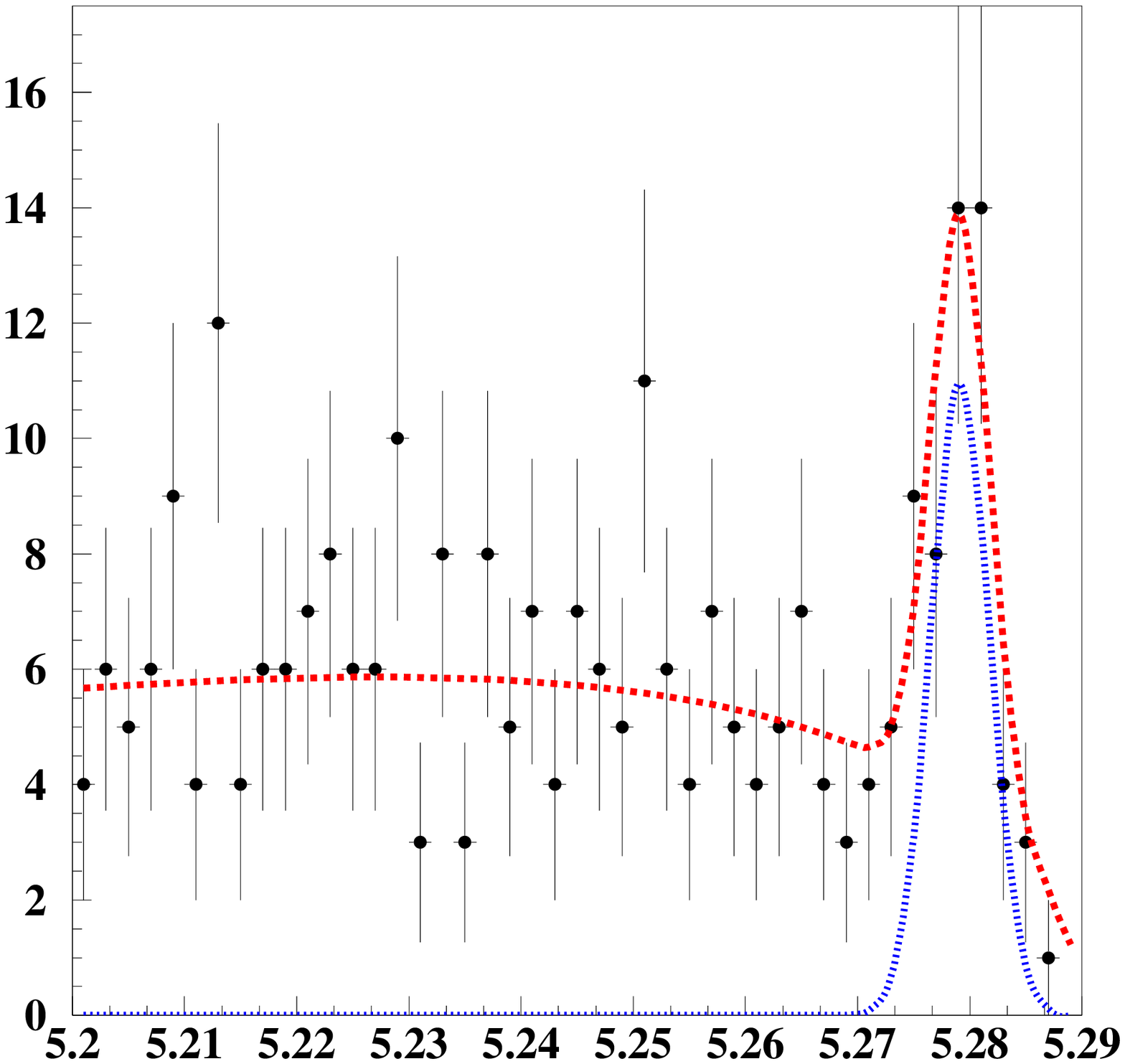}
\end{picture}
\end{minipage}\hfill
\begin{minipage}[b]{.32\linewidth}
\centering
\setlength{\unitlength}{1mm}
\begin{picture}(60,50)
\put(12,-3){\large $M(D_s^*)$[GeV/$c^{2}$]}
\put(-5,4)
{\rotatebox{90}{\large\bf $\frac{dN}{dM(D_s^*)\;\cdot\;(0.002\;\mathrm{GeV}/ c^2)}$}}
\includegraphics[height=4.5cm,width=5.4cm]{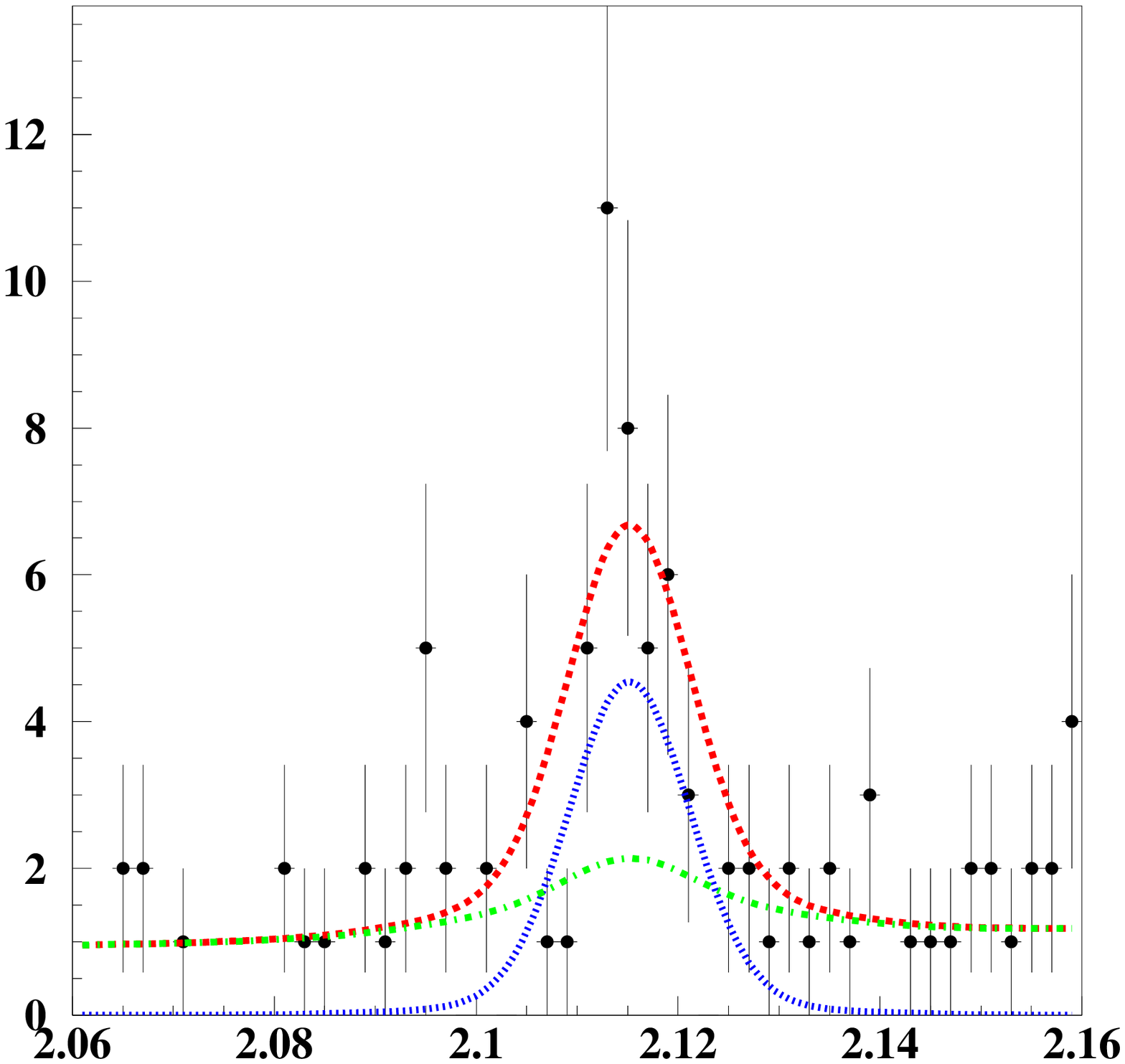}
\end{picture}
\end{minipage}\hfill

\caption{Distributions of $\Delta E$,  $M_\mathrm{bc}$ and $M(D_s^*) $
for {\bf (a)} $B^+\to D^{*-}_s(\to \phi\pi^-) K^+\pi^+$,
{\bf (b)} $B^+\to D^{*-}_s(\to K^{*0} K^-) K^+\pi^+$ and 
{\bf (c)} $B^+\to D^{*-}_s(\to K^0_S K^-) K^+\pi^+$ decays. 
For each distribution, $\Delta E$,  $M_\mathrm{bc}$ and $M(D_s^*) $ we select on
the signal region of the remaining two.
The red dashed curves show the results of the overall fit described in the text, the blue dashed curves correspond to signal components and the green dashed curves indicate the fitted background for $M(D_s^*) $.}



\label{FIG_DSGKAPI}
\end{figure}


The signal  yields are extracted using unbinned extended maximum-likelihood fits
to the  $(\Delta E, M_\mathrm{bc},
M(D_s^{(*)}))$ distributions of the selected candidate events.
 The likelihood function is given by:
\begin{equation}
  {\cal L} = \frac{1}{N!}(N_S+N_B)^N e^{-N_S-N_B} \,
\prod_{i=1}^{N} \left( \frac{N_S}{N_S + N_B}{\cal P}_S^i
+ \frac{N_B}{N_S + N_B}{\cal P}_B^i \right)\, ,
\end{equation}
where $i$ is the event identifier, $N$ is the total number of  events  in the fit 
and $N_S(N_B)$ is the number of signal and background events, respectively.
We use  Gaussian functions to parameterize the signal probability density function
 in $\Delta E$ and $M_{\mathrm bc}$
and a double Gaussian function with a common mean for the $M(D_s^{(*)})$ distribution:
\begin{equation*}
{\cal P}_S^i  = 
 {\cal G}(\Delta E^i;~\overline{\Delta E}, \sigma_{\Delta E}) \times
 {\cal G}(M^i_{\mathrm bc};~ m_B, \sigma_{M_{\mathrm bc}}) \times
\end{equation*}
\begin{equation}
\times
\left[  f^S_{D_s^{(*)}}~{\cal G}(M^i(D_s^{(*)});~m_{D_s^{(*)}}, \sigma^{(1)}_{D_s^{(*)}}) + 
(1-f^S_{D_s^{(*)}})~{\cal G}(M^i(D_s^{(*)});~m_{D_s^{(*)}}, \sigma^{(2)}_{D_s^{(*)}}) 
\right],
\end{equation}
where  $\overline{\Delta E}, m_B, m_{D_s^{(*)}}, \sigma_{\Delta E}, \sigma_{M_{\mathrm bc}},
 f^S_{D_s^{(*)}}, \sigma^{(1)}_{D_s^{(*)}}$ and $\sigma^{(2)}_{D_s^{(*)}}$ are fit parameters. The latter three, which describe the signal shape corresponding to the $m{D_s^{(*)}}$ distributions are fixed to the values obtained from the fit to the $B^+\to D_s^{(*)+}\overline{D}^0$ control channels. In addition, we use the $B^+\to D_s^{*+}\overline{D}^0$ data samples to fix the signal widths for $\Delta E$ and $M_{\mathrm bc}$ for the  $B^+\to D_s^{*-}K^+\pi^+$ decays.

The background is parameterized with a second-order polynomial ($p2$) in the $\Delta E$ distribution.
For the $M_{\mathrm bc}$ background distribution we choose
a parameterization that was first used by the ARGUS collaboration~\cite{ARGUS},
$f(M_{\mathrm bc},\zeta) \propto M_{\mathrm bc}\sqrt{1-(M_{\mathrm bc}/E_\mathrm{beam})^2}
e^{-\zeta(1-(M_{\mathrm bc}/E_\mathrm{beam})^2)}$, where $\zeta$ is a fit parameter.
Finally, the $M(D_s^{(*)})$ background distribution is described 
by the sum of a double Gaussian function and a second-order
polynomial: 

\begin{equation}
{\cal P}_B^i =
 p_2(\Delta E^i;~w_0,w_1,w_2)\times f(M^i_{\mathrm bc};~\zeta) \times
\end{equation}
\begin{equation*}
\times
\Big[ p_2(M^i(D_s^{(*)});~v_0,v_1,v_2) + 
\end{equation*}
\begin{equation*}
+ f^B_{D_s^{(*)}}  {\cal G}(M^i(D_s^{(*)});
~m_{D_s^{(*)}}, \sigma^{(1)}_{D_s^{(*)}}) + 
(1-f^B_{D_s^{(*)}}) {\cal G}(M^i(D_s^{(*)});~m_{D_s^{(*)}}, \sigma^{(2)}_{D_s^{(*)}})\Big].
\end{equation*}
The values of the variables $w_0,w_1,w_2, \zeta, v_0, v_1, v_2$ are determined in the fit, whereas the $f_{D_s^{(*)}}^B$ are fixed to the values resulting from the fits to the appropriate control channels.
Figures~\ref{FIG_DSKAPI} and~\ref{FIG_DSGKAPI} show the 
distributions of  $\Delta E, M_\mathrm{bc}$ and
$M(D_s^{(*)})$ together with the fits described above. 

 For decays containing a $K^*(892)^0$ meson a small correction was applied to the signal yields obtained from the fit. The $K^*(892)^0$ mass sidebands (0.746-0.796) GeV/c$^2$ and
  (0.996-1.046) GeV/c$^2$ were fitted and a significant background contributing to the signal yields was found for the  $B^+\to D_s^-(\to K^{*0} K^-) K^+\pi^+$ and $B^+\to D_s^+(\to K^{*0} K^-) \overline{D^0}$ channels. Final signal yields were obtained by subtracting these contributions from the nominal fit values.

The signal yields together with statistical significances are listed in Table~\ref{RESULTS}. The significance is defined as $\sqrt{-2 {\rm ln}({\cal L}_0/{\cal L}_\mathrm{max})}$, where
${\cal L}_\mathrm{max}$ (${\cal L}_0$) denotes the  maximum likelihood with the signal yield at its
nominal value (fixed to zero).

The reconstruction efficiencies, determined using MC samples of
$e^+e^-\to \Upsilon(4S)\to B^+ B^-$ decays, are listed in Table~\ref{RESULTS}. 
This table also contains the values obtained for the branching fractions of the decays $B^+ \rightarrow
D_s^{(*)-} K^+\pi^+$ and $B^+ \rightarrow D_s^{(*)+}\overline{D^0}$. The last error (Table~\ref{RESULTS}) is due to uncertainties in the branching fractions for the decays of intermediate
particles, predominantly the $D^{(*)}_s$~\cite{PDG}. The systematic uncertainties are evaluated only for
the three-body $B^+\to D_s^{(*)-}K^+\pi^+$ decays.
We find branching fractions for the $B^+ \rightarrow D_s^{(*)+}\overline{D^0}$ control samples in agreement with 
world averages~\cite{PDG}.

\begin{table}[bt]
\begin{center}
\caption{Signal yields, reconstruction efficiencies, branching fractions
 and statistical significances  for 
$B^+\to D_s^{(*)-} K^+\pi^+$ and $B^+\to D_s^{(*)+}
K^+\pi^-$ decays.}
\vspace*{0.5ex}
\begin{tabular}{lcccc}
\hline
Decay &  Signal  & Efficiency & Statistical        & Branching              \\
      &  yield   & [\%]       & Signif. [$\sigma$] & fraction [$(10^{-4})$] \\
\hline
$B^+\to D_s^-(\to \phi\pi^-) K^+\pi^+$ & $306.0^{+19.7}_{-19.1}$ & $11.25 \pm 0.11$ & $31.5$ & 
$1.90^{+0.12~+0.18}_{-0.12~-0.18} \pm 0.29$ \\
$B^+\to D_s^-(\to K^{*0} K^-) K^+\pi^+$ & $281.7^{+24.7}_{-23.6}$ & $8.55 \pm 0.10$ & $26.5$ &
$1.93^{+0.17~+0.19}_{-0.16~-0.19} \pm 0.30$ \\
$B^+\to D_s^-(\to K^0_S K^-) K^+\pi^+$ & $179.4^{+16.7}_{-16.0}$ & $12.82 \pm 0.19$ & $20.4$ &
$2.06^{+0.19~+0.25}_{-0.18~-0.26} \pm 0.13$ \\
$B^+\to D_s^{*-}(\to \phi\pi^-) K^+\pi^+$ & $59.0^{+9.3}_{-8.6}$ & $3.00 \pm 0.06$ & $11.0$ &
$1.46^{+0.23~+0.18}_{-0.21~-0.19} \pm 0.22$ \\
$B^+\to D_s^{*-}(\to K^{*0} K^-) K^+\pi^+$ & $61.7^{+10.6}_{-9.8}$ & $2.65 \pm 0.06$ & $9.3$ &
$1.45^{+0.25~+0.18}_{-0.23~-0.19} \pm 0.22$ \\
$B^+\to D_s^{*-}(\to K^0_S K^-) K^+\pi^+$ & $35.7^{+7.7}_{-6.9}$ & $3.67 \pm 0.11$ & $8.0$ &
$1.53^{+0.33~+0.24}_{-0.30~-0.22} \pm 0.09$ \\
\hline
$B^+\to D_s^+(\to \phi\pi^+) \overline{D^0}$ & $597.4^{+25.0}_{-24.3}$ & $13.03 \pm 0.12$ & $56.8$ &
$82.31^{+3.45}_{-3.35}\pm 12.50$ \\
$B^+\to D_s^+(\to \overline{K^{*0}} K^+) \overline{D^0}$ & $512.6^{+26.2}_{-25.3}$ & $9.21 \pm 0.10$ & $53.3$ &
$83.80^{+4.28}_{-4.14}\pm 12.94$ \\
$B^+\to D_s^+(\to K^0_S K^+) \overline{D^0}$ & $294.5^{+17.8}_{-17.2}$ & $14.22 \pm 0.20$ & $38.9$ &
$78.61^{+4.74}_{-4.56}\pm 4.86$ \\
$B^+\to D_s^{*+}(\to \phi\pi^+) \overline{D^0}$ & $150.2^{+15.7}_{-14.8}$ & $3.97 \pm 0.07$ & $19.0$ &
$72.15^{+7.52}_{-7.12}\pm 10.98$ \\
$B^+\to D_s^{*+}(\to \overline{K^{*0}} K^+) \overline{D^0}$ & $151.9^{+15.1}_{-14.3}$ & $3.09 \pm 0.06$ & $20.8$ &
$78.68^{+7.83}_{-7.43}\pm 12.16$ \\
$B^+\to D_s^{*+}(\to K^0_S K^+) \overline{D^0}$ & $95.3^{+12.4}_{-11.6}$ & $4.40 \pm 0.12$ & $15.0$ &
$87.27^{+11.32}_{-10.63}\pm 5.43$ \\
\hline
\end{tabular}
\label{RESULTS}
\end{center}
\end{table}

\begin{table}[hbt]
\begin{center}
\caption{Systematic uncertainties on the branching fractions for $B^+\to D_s^{(*)-} K^+\pi^+$ decay modes, given in percent.} 
\vspace*{0.5ex}
\begin{tabular}{lcccccc}
\hline
\multicolumn{1}{c}{Source} &  
\multicolumn{3}{c}{$D_s^-$ final state} &
\multicolumn{3}{c}{$D_s^{*-}$ final state} \\
\cline{2-7} 
\multicolumn{1}{c}{}             & 
\multicolumn{1}{c}{$\phi \pi^-$} & 
\multicolumn{1}{c}{$K^{*0}K^-$}  &
\multicolumn{1}{c}{$K^0_S K^-$}  &
\multicolumn{1}{c}{$\phi \pi^-\gamma$} & 
\multicolumn{1}{c}{$K^{*0}K^-\gamma$}  &
\multicolumn{1}{c}{$K^0_S K^-\gamma$}  \\ 
\hline
(a) Tracking                       & 5 & 5 & 5 & 5 & 5 & 5 \\
(b) Hadron identification          & 5 & 5 & 5 & 5 & 5 & 5 \\
(c) $K^0_S$ reconstruction         & - & - & 4.5 & - & - & 4.5 \\
(d) Photon reconstruction          & - & - & - & 5 & 5 & 5 \\
(e) Uncertainty in N($B\overline{B}$) & 1.4 & 1.4 & 1.4 & 1.4 & 1.4 & 1.4 \\
(f) Selection procedure            & 3.7 & 3.7 & 3.7 & 3.7 & 3.7 & 3.7 \\
(g) Size of candidate region       & 0.6 & 1.1 & 1.2 & 0.3 & 0.2 & 1.5 \\
(h) Signal shape                   & $~^{+1.2}_{-1.3}$ & $~^{+3.4}_{-3.4}$ & $~^{+5.6}_{-6.3}$ & $~^{+5.6}_{-6.7}$ & $~^{+6.9}_{-7.4}$ & $~^{+9.7}_{-7.8}$ \\
(i) MC statistics                  & 5.1 & 4.1 & 5.2 & 5.4 & 4.5 & 5.8 \\
\hline
Total  & $~^{+9.6}_{-9.7}$ & $~^{+9.8}_{-9.8}$ & $~^{+12.1}_{-12.4}$ & $~^{+12.3}_{-12.8}$ & $~^{+12.6}_{-12.9}$ & $~^{+15.5}_{-14.4}$\\
\hline
\end{tabular}
\label{SYST}
\end{center}
\end{table}

Systematic uncertainties are listed in Table~\ref{SYST}.
The contribution (f) due to the selection procedure is dominated by the $R_2$ requirement. This
uncertainty is estimated conservatively as the maximum variation of the 
efficiency-corrected signal yield, when the $R_2$ selection value is varied 
over a wide range (values between 0.25 and 0.55).
The uncertainty (g) due to the fit range is determined by varying 
the candidate region. 
To evaluate the contribution (h) we repeat the fits varying the shape parameters by $\pm 1\sigma$.
The uncertainty (i) is estimated as the statistical error in the selection efficiency,  increased conservatively by a factor obtained from the difference between the value of the branching fraction for the appropriate control channel and the generated branching fraction.
The overall
systematic error is obtained by summing these contributions in
quadrature. 

The average branching  fractions for the decays
$B^+\to D_s^- K^+\pi^+$ and
$B^+\to D_s^{*-} K^+\pi^+$ are determined from a simultaneous fit to the data containing events from all three $D_s$ decay modes. Here, the systematic uncertainties are calculated as in the individual channels (Table~\ref{SYST}).


In summary, the following branching fractions are determined:
\begin{equation}
{\cal B}(B^+\to D_s^- K^+\pi^+) = (1.94^{+0.09}_{-0.08} ({\mathrm stat})^{~+0.20}_{~-0.20} ({\mathrm syst})   \pm 0.17 ({\mathrm {\cal B}_{int})})\times 10^{-4}~~
\end{equation}
\begin{equation}
{\cal B}(B^+\to D_s^{*-} K^+\pi^+) = (1.47^{+0.15}_{-0.14} ({\mathrm stat})^{~+0.19}_{~-0.19} ({\mathrm syst})   \pm 0.13 ({\mathrm {\cal B}_{int})})\times 10^{-4}.~~
\end{equation}
These branching fractions are compatible with the values reported by
the BaBar collaboration~\cite{BABAR_DSKAPI}. 
 
\begin{figure}[tbh]
\begin{minipage}[b]{.46\linewidth}
\centering
\setlength{\unitlength}{1mm}
\begin{picture}(95,42)
\put(60,39){\bf (a)}
\put(15,-2){\large $M(D_s K)$[GeV/$c^2$]}
\put(-3,5){\rotatebox{90}{\bf $\frac{dN}{d(M(D_s K))\;\cdot\;(0.067\;\mathrm{GeV}/c^2)} $}}
\includegraphics[width=8cm,height=5.0cm]{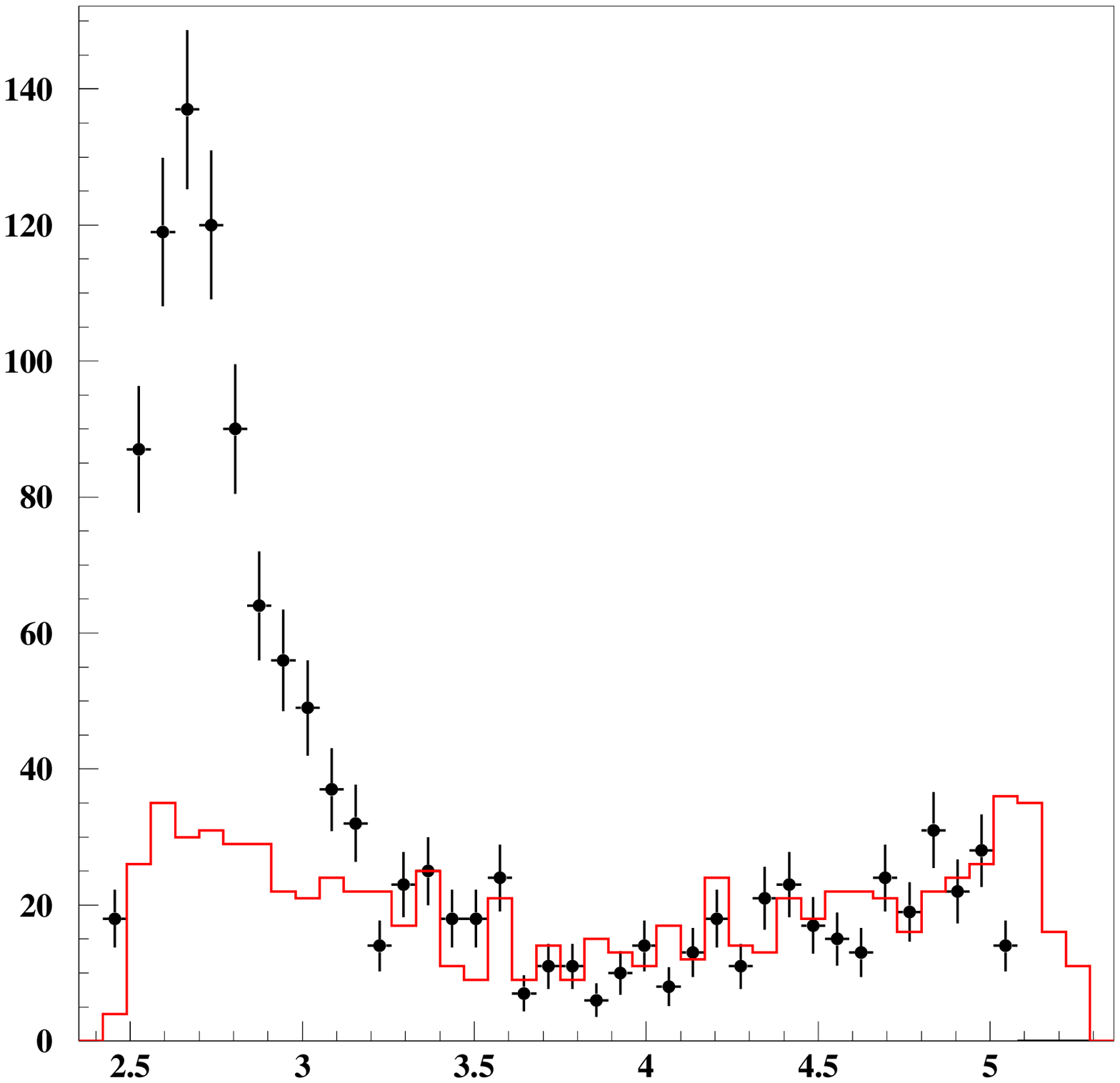}
\end{picture}
\end{minipage}\hfill
\begin{minipage}[b]{.46\linewidth}
\centering
\setlength{\unitlength}{1mm}
\begin{picture}(95,42)
\put(60,39){\bf (b)}
\put(15,-2){\large $M(D_s^* K)$[GeV/$c^2$]}
\put(-3,5){\rotatebox{90}{\bf $\frac{dN}{d(M(D_s^* K))\;\cdot\;(0.067\;\mathrm{GeV}/c^2)} $}}
\includegraphics[width=8cm,height=5.0cm]{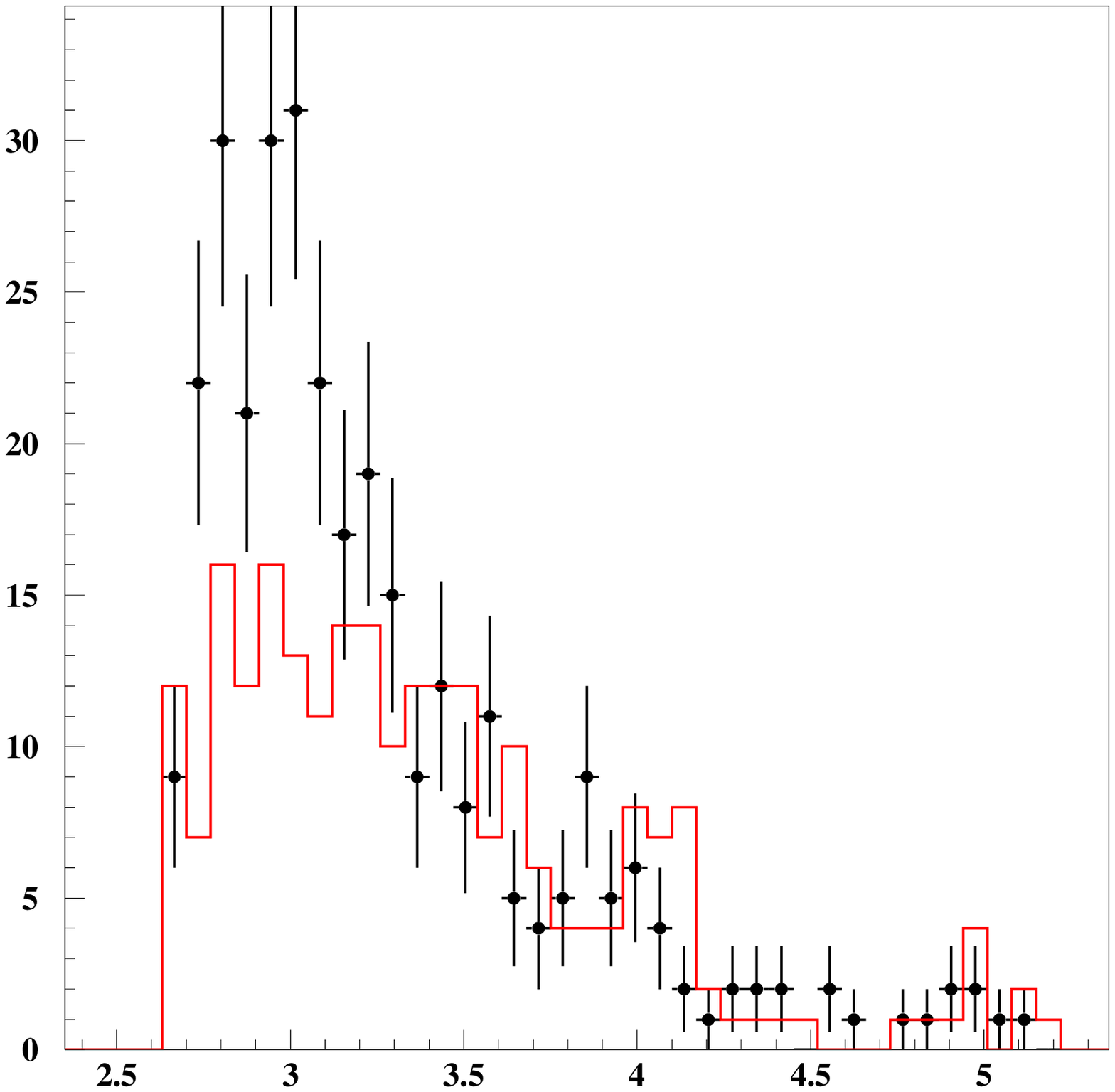}
\end{picture}
\end{minipage}
\caption{ The invariant mass distributions of {\bf (a)} $D_s^-K^+$ 
 for the decay $B^+\to D_s^- K^+\pi^+$ 
and {\bf (b)} of   $D_s^{*-}K^+$ 
 for  $B^+\to D_s^{*-} K^+\pi^+$ 
corresponding to the signal regions described in the text.
The histograms show the background contributions corresponding 
to $\Delta E > 0.06$ GeV.
}
\label{DSKPAIR}
\end{figure}

The invariant mass distributions of the $D_s^{(*)-}K^+$ subsystem
are incompatible with those expected for three-body phase space production and exhibit
strong enhancements around 2.7 GeV/$c^2$~(see Fig.~\ref{DSKPAIR}). These features may be explained by the production of charm resonances with masses below $D_s^{(*)-}K^+$ threshold~\cite{THEOR}.

We thank the KEKB group for the excellent operation of the
accelerator, the KEK cryogenics group for the efficient
operation of the solenoid, and the KEK computer group and
the National Institute of Informatics for valuable computing
and SINET3 network support.  We acknowledge support from
the Ministry of Education, Culture, Sports, Science, and
Technology (MEXT) of Japan, the Japan Society for the 
Promotion of Science (JSPS), and the Tau-Lepton Physics 
Research Center of Nagoya University; 
the Australian Research Council and the Australian 
Department of Industry, Innovation, Science and Research;
the National Natural Science Foundation of China under
contract No.~10575109, 10775142, 10875115 and 10825524; 
the Department of Science and Technology of India; 
the BK21 program of the Ministry of Education of Korea, 
the CHEP src program and Basic Research program (grant 
No. R01-2008-000-10477-0) of the 
Korea Science and Engineering Foundation;
the Polish Ministry of Science and Higher Education;
the Ministry of Education and Science of the Russian
Federation and the Russian Federal Agency for Atomic Energy;
the Slovenian Research Agency;  the Swiss
National Science Foundation; the National Science Council
and the Ministry of Education of Taiwan; and the U.S.\
Department of Energy.
This work is supported by a Grant-in-Aid from MEXT for 
Science Research in a Priority Area ("New Development of 
Flavor Physics"), and from JSPS for Creative Scientific 
Research ("Evolution of Tau-lepton Physics").

\newpage




\begin{thebibliography}{99}

\bibitem{FOOT} 
  Throughout this paper, the inclusion of the charge-conjugate decay mode is implied unless otherwise stated.

\bibitem{BABAR_DSKAPI} 
  B.~Aubert \textit{et al.} (BaBar Collaboration), Phys. Rev. Lett.  {\bf 100}, 171803 (2008).


\bibitem{THEOR} 
  O.~Antipin and G.~Valencia, Phys. Lett. B \textbf{647}, 164 (2007).

\bibitem{KEKB}
  S.~Kurokawa and E.~Kikutani, Nucl. Instr. and Meth. A \textbf{499}, 1 (2003),
  and other papers included in this volume.

\bibitem{BELLE}  
  A.~Abashian \textit{et al.} (Belle Collaboration), Nucl. Instr. and Meth. A \textbf{479}, 117 (2002).

\bibitem{SVD} 
  Z.~Natkaniec \textit{et al.} (Belle SVD2 Group), Nucl. Instr. and Meth. A \textbf{560}, 1 (2006).

\bibitem{PDG}  
  C. Amsler et al., Phys. Lett. B \textbf{667}, 1 (2008).


\bibitem{GEANT}  
  R.~Brun \textit{et al.} GEANT 3.21, CERN REPORT DD/EE/84-1, 1984.

\bibitem{FOX} 
  G.~C.~Fox and S.~Wolfram, Phys. Rev. Lett. \textbf{41}, 1581 (1978).

\bibitem{ARGUS} 
  H.~Albrecht \textit{et al.} (ARGUS Collaboration), Phys. Lett. B \textbf{241}, 278 (1990).



\end{thebibliography}
\end{document}